                        \newif\ifboyscout                         
                        \boyscoutfalse 
                        \newif\ifpreparepdf                       
                        \preparepdftrue 

        \ifboyscout
\documentclass[onecollarge]{svjour3}       
        \else
\documentclass[onecollarge,final]{svjour3}       
        \fi

\ifpreparepdf
    \usepackage{color} 
    \usepackage[colorlinks]{hyperref} 
\else 
\fi
    \usepackage{graphicx,amssymb,amsbsy,amsmath,amsfonts,longtable,calc,array,floatrow,setspace}
    \usepackage{ifthen}
    \usepackage[percent]{overpic}
    \usepackage{hyperref}

\ifpreparepdf 

\else  

\fi
\graphicspath{{figs/}}  

\newcommand{\sspTW}{\ensuremath{\ssp_{tw}}}
\newcommand{\sspRPO}{\ensuremath{\ssp_{rp}}}       
\newcommand{\sspEQ}{\ensuremath{\ssp_{eq}}}
\newcommand{\sspPPO}{\ensuremath{\ssp_{pp}}}       
\newcommand{\sspRedPPO}{\ensuremath{\sspRed_{pp}}} 

\newcommand{\Lg}{\ensuremath{T}}   
\newcommand{\matrixRep}{\ensuremath{{D}}}  
\newcommand{\LieEl}{\ensuremath{g}}  

\newcommand{\rf}     [1] {\,\cite{#1}}
\newcommand{\refref} [1] {ref.~\cite{#1}}
\newcommand{\refRef} [1] {Ref.~\cite{#1}}
\newcommand{\refrefs}[1] {refs.~\cite{#1}}
\newcommand{\refRefs}[1] {Refs.~\cite{#1}}
\newcommand{\refeq}  [1] {(\ref{#1})}
\newcommand{\reffig} [1] {fig. \ref{#1}}
\newcommand{\refFig} [1] {Fig. \ref{#1}}
\newcommand{\reftab} [1] {table \ref{#1}}

\newcommand{\refsect} [1] {section \ref{#1}}

\newcommand{\refappe}[1] {appendix \ref{#1}}

\newcommand{\beq}{\begin{equation}}
\newcommand{\continue}{\nonumber \\ }
\newcommand{\nnu}{\nonumber}
\newcommand{\eeq}{\end{equation}}
\newcommand{\ee}[1] {\label{#1} \end{equation}}
\newcommand{\bea}{\begin{eqnarray}}

\newcommand{\eea}{\end{eqnarray}}

\newcommand{\ie}{{i.e.}}            
\newcommand{\eg}{{e.g.}}
\newcommand{\etal}{{\em et al.}}    

\newcommand{\statesp}{state space}

\newcommand{\dmn}{-dimensional}  

\newcommand{\jacobianM}{Jacobian matrix}  
\newcommand{\jacobianMs}{Jacobian matrices}   
\newcommand{\ExpaEig}{\Lambda}

\newcommand{\zeit}{\ensuremath{\tau}}  
\newcommand\period[1]{{\ensuremath{T_{#1}}}}         

\renewcommand\Im{\ensuremath{{\rm Im}\,}}
\renewcommand\Re{\ensuremath{{\rm Re}\,}}

\newcommand{\po}{periodic orbit}

\newcommand{\rpo}{rela\-ti\-ve periodic orbit}
\newcommand{\Rpo}{Rela\-ti\-ve periodic orbit}
\newcommand{\ppo}{pre-periodic orbit}

\newcommand{\eqv}{equi\-lib\-rium}
\newcommand{\Eqv}{Equi\-lib\-rium}
\newcommand{\eqva}{equi\-lib\-ria}

\newcommand{\reqv}{rela\-ti\-ve equi\-lib\-rium}
\newcommand{\Reqv}{Rela\-ti\-ve equi\-lib\-rium}
\newcommand{\reqva}{rela\-ti\-ve equi\-lib\-ria}

\newcommand{\bseq}{\begin{subequations}}
\newcommand{\eseq}{\end{subequations}}

\newcommand{\NS}{Navier-Stokes}

\newcommand{\NSe}{Navier-Stokes equations}

\newcommand{\slicePlane}{slice hyperplane}

\newcommand{\Poincare}{Poincar\'e }
\newcommand{\PoincSec}{Poincar\'e section}

\newcommand{\slice}{slice}

\newcommand{\mslices}{method of slices}

\newcommand{\sspRed}{\ensuremath{\hat{\ssp}}}    
\newcommand{\velRed}{\ensuremath{\hat{\vel}}}    

\newcommand{\phaseVel}{phase velocity}      

\newcommand{\PoincS}{\ensuremath{{\cal P}}}  

\newcommand{\On}[1]{\ensuremath{\textrm{O}(#1)}}
\newcommand{\SOn}[1]{\ensuremath{\textrm{SO}(#1)}}         
\newcommand{\Ztwo}{\ensuremath{\textrm{C}_2}}                

\newcommand{\KS}{Ku\-ra\-mo\-to-Siva\-shin\-sky}
\newcommand{\KSe}{Ku\-ra\-mo\-to-Siva\-shin\-sky equation}

\newcommand{\eigExp}[1][]{
     \ifthenelse{\equal{#1}{}}{\ensuremath{\lambda}}{\ensuremath{\lambda^{(#1)}}}}
\newcommand{\eigRe}[1][]{
     \ifthenelse{\equal{#1}{}}{\ensuremath{\mu}}{\ensuremath{\mu^{(#1)}}}}
\newcommand{\eigIm}[1][]{
     \ifthenelse{\equal{#1}{}}{\ensuremath{\omega}}{\ensuremath{\omega^{(#1)}}}}

\newcommand\flow[2]{{f^{#1}(#2)}}
\newcommand{\vel}{\ensuremath{v}}   

\newcommand{\ssp}{\ensuremath{a}}                
\newcommand{\fFslice}{first Fourier mode slice}

\newcommand{\inprod}[2]{\langle #1 ,\, #2 \rangle}  

\newcommand{\jMps}{\ensuremath{J}}   

\newcommand{\conf}{\ensuremath{x}} 

\newcommand{\Fu}{\tilde{u}}

\newcommand{\jMpsRed}{\ensuremath{\hat{\jMps}}}   
\newcommand{\jMpsRefRed}{\ensuremath{\tilde{\jMps}}}   
\newcommand\flowRed[2]{{\hat{f}^{#1}(#2)}}
\newcommand\flowRefRed[2]{{\tilde{f}^{#1}(#2)}}

\newcommand{\sspRefRed}{\ensuremath{\tilde{\ssp}}} 
\newcommand{\velRefRed}{\ensuremath{\tilde{\vel}}} 
\newcommand{\matrixRepRed}{\ensuremath{\hat{\matrixRep}}}  
\newcommand{\primeOrb}[1]{\ensuremath{p_#1}}
\newcommand{\VRed}{\ensuremath{\hat{V}}}  
\newcommand{\VRefRed}{\ensuremath{\tilde{V}}}  
\newcommand{\ProjPsect}{\mathbb{P}}
\newcommand{\matId}{\ensuremath{{\bf 1}}}      

\journalname{Journal of Statistical Physics}

\begin{document}
\titlerunning{Unstable manifolds of relative periodic orbits}
\title{
Unstable manifolds of relative periodic orbits
in the symmetry-reduced state space of the Kuramoto-Sivashinsky system
\thanks{
This work was supported by the family of
late G. Robinson, Jr. and NSF grant DMS-1211827.}
}

\author{Nazmi Burak Budanur \\
        Predrag Cvitanovi\'c
}


\institute{Nazmi Burak Budanur \at
           Institute of Science and Technology (IST),
           3400 Klosterneuburg, Austria. 
              \email{burak.budanur@ist.ac.at}           
           \and
           Predrag Cvitanovi\'c\at
           Center for Nonlinear Science, School of Physics,
           Georgia Institute of Technology,
           Atlanta, GA 30332-0430, USA.
}

\date{Received: August 20, 2016 / Accepted: date}

\maketitle

\begin{abstract}
Systems such as fluid flows in channels and pipes or the complex
Ginz\-burg-Landau system, defined over periodic domains, exhibit both
continuous symmetries, translational and rotational, as well as discrete
symmetries under spatial reflections or complex conjugation. The
simplest, and very common symmetry of this type is the equivariance of
the defining equations under the orthogonal group O(2). We formulate a
novel symmetry reduction scheme for such systems by combining the method
of slices with invariant polynomial methods, and show how it works by
applying it to the Kuramoto-Sivashinsky system in one spatial dimension.
As an example, we track a relative periodic orbit through a sequence of
bifurcations to the onset of chaos. Within the symmetry-reduced state
space we are able to compute and visualize the unstable manifolds of
relative periodic orbits, their torus bifurcations, a transition to chaos
via torus breakdown, and heteroclinic connections between various
relative periodic orbits. It would be very hard to carry through such
analysis in the full state space, without a symmetry reduction such as
the one we present here.
\end{abstract}

\newpage
\section{Introduction}

\begin{quotation}
``\ldots of course, the motion of the system tends to
move away from repellers. Nonetheless a repeller might be
important because, for example, it might describe a separatrix
that serves to divide two different attractors or two different
types of motion.'' \emph{Kadanoff and Tang\rf{KT84}}.
\end{quotation}

The 1984 Kadanoff and Tang investigation of strange repellers was prescient
in two ways. First, at the time it was not obvious why  anyone would care
about ``repellers,'' as their dynamics would be transient. Today, much of the
research in turbulence focuses on repellers. In particular, significant
effort is invested in understanding the state space regions of shear-driven
fluid flows that separate laminar and turbulent 
regimes\rf{TI03,SYE05,SchEckYor07,duguet07,AvMeRoHo13}, 
and these ``separatrices'' indeed often
appear to be strange repellers.
Kadanoff and Tang's study was quantitative, and modest by today's standards:
they computed escape rates for a family of 3-dimensional mappings in terms of
their unstable \po s (`repulsive cycles'), while today corresponding
computations are carried out for very high-dimensional ($\sim$100,000
computational degrees of freedom), numerically accurate discretizetions of
Navier-Stokes flows\rf{GHCW07,channelflow,openpipeflow}. In light of the
heuristic nature of their investigation, their second insight was remarkable:
they were the first to posit the {\em exact} weight for the contribution of an
unstable periodic orbit $p$ to an average computed over a strange
repeller (or attractor):
\[
    {1}
    /
    {\left|\det \left( {\bf 1}-\jMps_p(x) \right)\right| }
\]    
(here $\jMps_p(x)$ is the Jacobian matrix of linearized flow, computed along
the orbit of a periodic point $x$). While, at the time, they were aware only
of Bowen's (1975) work\rf{bowen}, today this formula is a cornestone of the
modern periodic orbit theory of chaos in deterministic flows\rf{DasBuch},
based on zeta functions of Smale (1967)\rf{smale}, Gutzwiller
(1969)\rf{gutzwiller71}, Ruelle (1976)\rf{Ruelle76a,Ruelle76} and their
cycle expansions (1987)\rf{inv,AACI,AACII,CBook:appendHist}. Much has
happened since -- in particular, the formulas of periodic orbit theory for
3-dimensional dynamics that they had formulated in 1983 are today at the core
of the challenge very dear to Kadanoff, a dynamical theory of
turbulence\rf{Christiansen97,GHCW07}. For that to work, many extra moving
parts come into play. We have learned that the convergence of cycle
expansions relies heavily on the flow topology and the associated symbolic
dynamics, and that understanding the geometry of flows in the state space is
the first step towards extending periodic orbit theory to systems of high or
infinite dimensions, such as fluid flows. It turns out that taking care of
the symmetries of a nonlinear fluid flow is also a difficult problem. While
one can visualize dynamics in 2 or 3 dimensions, the \statesp\ of these flows
is high-dimensional, and symmetries -both continuous and discrete- complicate
the flow geometry as each solution comes along with all of its symmetry
copies. In this contribution to Leo Kadanoff memorial volume, we develop new
tools for investigating geometries of flows with symmetries, and illustrate
their utility by applying them to a spatiotemporally chaotic \KS\ system.

Originally derived as a simplification of the complex Ginzburg-Landau
equation\rf{KurTsu76} and in study of flame fronts\rf{siv}, the \KS\
is perhaps the simplest spatially extended dynamical system that exhibits
spatiotemporal chaos. Similar in form to the \NSe, but much easier
computationally, the \KS\ partial differential equation (PDE) is a
convenient sandbox for developing intuition about
turbulence\rf{Holmes96}. As for the \NS, a state of the \KS\ system is
usually visualized by its shape over configuration space
(such as states shown in \reffig{f-ksconf}). However, the function space
of allowable PDE fields is an infinite-dimensional state space, with the
instantaneous state of the field a point in this space. In spite of the
\statesp\ being of high (and even infinite) dimension, evolution of the
flow can be visualized, as generic trajectories are 1-dimensional
curves, and numerically exact solutions such as equilibria and periodic
orbits are points or closed loops, in any \statesp\ projection.
There are many choices of a ``\statesp.'' Usually one starts out with the
most immediate one: computational elements used in a
finite-dimensional discretization of the PDE studied. As the \KS\ system
in one space dimension, with periodic boundary condition, is equivariant
under continuous translations and a reflection, for the case at hand the
natural choice is a Fourier basis, truncated to a desired numerical
accuracy. This is still a high-dimensional space: in numerical work
performed here, $30$-dimensional. For effective visualizations, one
thus needs to carefully pick dynamically intrinsic coordinate frames, and
projections on them\rf{SCD07,GHCW07}.

Such dynamical systems visualisations of turbulent flows, complementary
to the traditional spatio-temporal visualizations, offer invaluable
insights into the totality of possible motions of a turbulent fluid.
However, symmetries, and especially continuous symmetries, such as
equvariance of the defining equations under spatial translations, tend to
obscure the \statesp\ geometry of the system by their preference for
higher-dimensional invariant $N$-tori solutions, such as \reqva\ and \rpo s.

In order to avoid dealing with such effects of continuous symmetry, a
number of papers%
\rf{Christiansen97,RCMR04,ReCi05,lanCvit07,ReChMi07} study the \KSe\
within the flow-invariant subspace of solutions symmetric under
reflection. However, such
restrictions to flow-invariant subspaces miss the physics of the problem:
any symmetry invariant subspace is of zero measure in the full \statesp,
so a generic turbulent trajectory explores the \statesp\ \emph{outside} of
it. Lacking continuous-symmetry reduction schemes, earlier papers
on the geometry of the \KS\ flow in the full \statesp%
\rf{ksgreene88,AGHks89,KNSks90,SCD07} were restricted to the study of
the smallest invariant structures: \eqva, their stable/unstable
manifolds, their heteroclinic connections, and their bifurcations under
variations of the domain size.

Stationary solutions are important for understanding the \statesp\
geometry of a chaotic attractor, as their stable manifolds typically
set the boundaries of the strange set. The Lorenz attractor is the best
known example\rf{Williams79} and Gibson \etal\rf{GHCW07} visualizations
for the plane Couette flow are so far the highest-dimensional setting,
where this claim appears to hold. In this paper we turn our attention to
(relative) \po s, which --unlike unstable \eqva-- are embedded within the
strange set, and are expected to capture physical properties of an
ergodic flow.
\refRefs{Christiansen97,lanCvit07}, restricted to the
reflection-invariant subspace of the \KS\ flow, have succeeded in
constructing symbolic dynamics for several system sizes. In these
examples, short \po s have real Floquet multipliers, with very thin
unstable manifolds, around which the longer \po s are organized by means
of nearly 1\dmn\ \Poincare\ return maps.

In this paper we study the unstable manifolds of \rpo s of \KS\ system in
full \statesp, with no symmetry restrictions. In contrast to the
flow-invariant subspace considered in \refrefs{Christiansen97,lanCvit07},
the shortest \rpo\ of the full system that is stable for small system
sizes ($L < 21.22$) has a complex leading Floquet multiplier. This
renders the associated unstable manifold 2-dimensional. Elimination of
the marginal directions, the space and time translation symmetries,
by a `slice' and a {\PoincSec} conditions, together with a novel
reduction of the spatial reflection symmetry, enables us to study here
this 2-dimensional unstable manifold. We compute and visualize the
unstable manifold of the shortest \po\ as we increase the system size
towards the system's transition to chaos.

Summary of our findings is
as follows:
At the system size $L \approx 21.22$, the shortest \po\ undergoes a torus
bifurcation\rf{hopf42} (also sometimes referred to as
the Neimark-Sacker bifurcation\rf{Neimark59,Sacker65},
if the flow is studied in a {\PoincSec}), which gives birth to a
stable 2-torus. As the system size is increased, this torus first
goes unstable, and is eventually destroyed by
the bifurcation into  stable and unstable pair of period-$3$ orbits, to which the
unstable manifold of the parent orbit is heteroclinically connected.
As the system size is increased further, the stable period-$3$ orbit goes
unstable, then disappears, and the dynamics becomes chaotic.
Upon a further increase of the system size, the unstable period $3$ orbit
undergoes a symmetry-breaking bifurcation, which introduces richer
dynamics as the associated unstable manifold has connections to both
drifting (relative) and non-drifting \po s.

We begin by a short review of the \KS\ system in the next section, and
review continuous symmetry reduction by {\fFslice} method in
\refsect{s-SymmRedCont}. The main innovation introduced in this paper is the
invariant polynomial discrete symmetry reduction method described in
\refsect{s-SymmRedDiscr}. The new symmetry reduction method is applied to and
tested on the \KS\ system in \refsect{s-UnstMan}, where the method
makes it possible to track the
evolution of the \po s unstable manifolds through the system's
transition to chaos. We discuss the implications of our results and
possible future directions in \refsect{s-Discuss}.

\section{\KS\ system and its symmetries}
\label{s-kse}

\begin{figure}[h]
    \centering
    \begin{overpic}[height=0.30\textwidth]{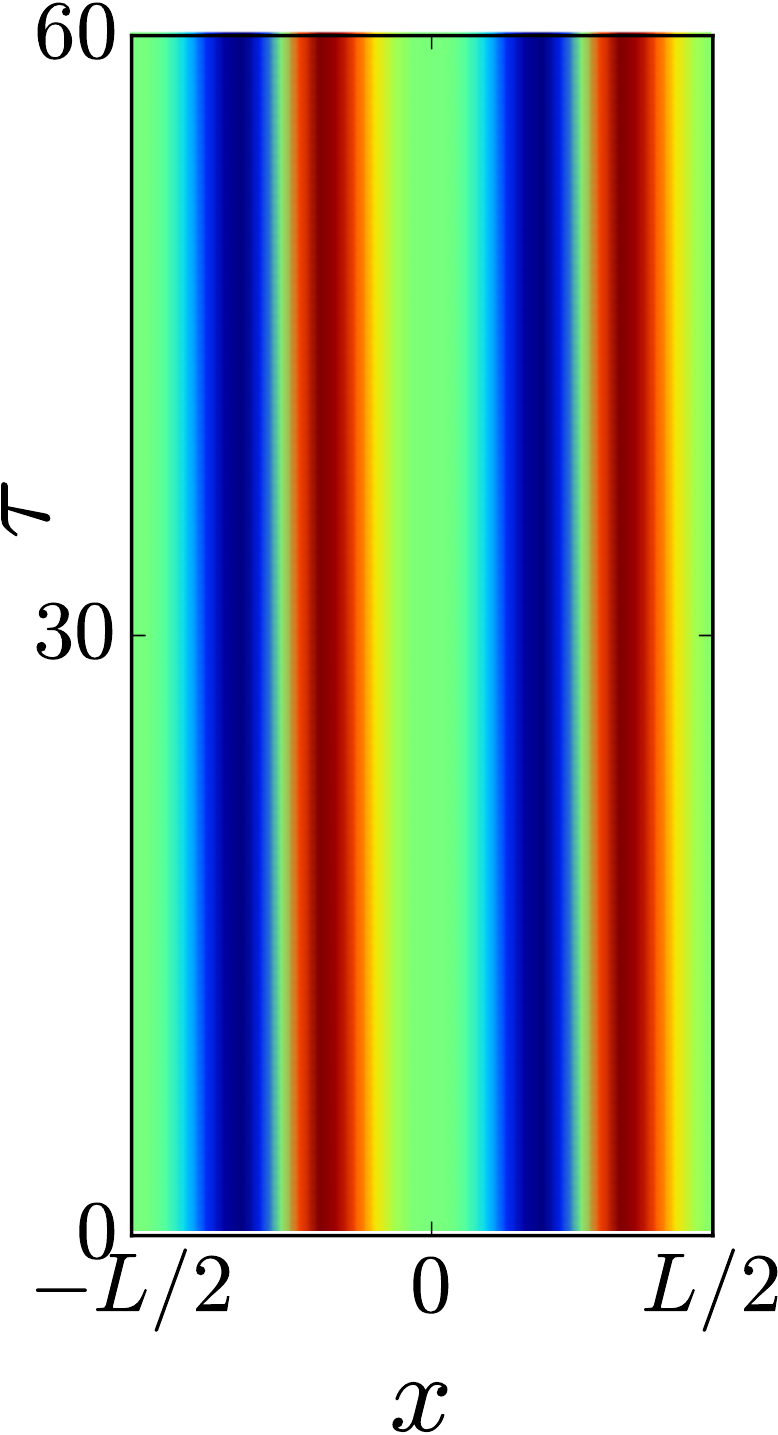}
        \put (-0.75,-1) {(a)}
    \end{overpic} \quad
    \begin{overpic}[height=0.30\textwidth]{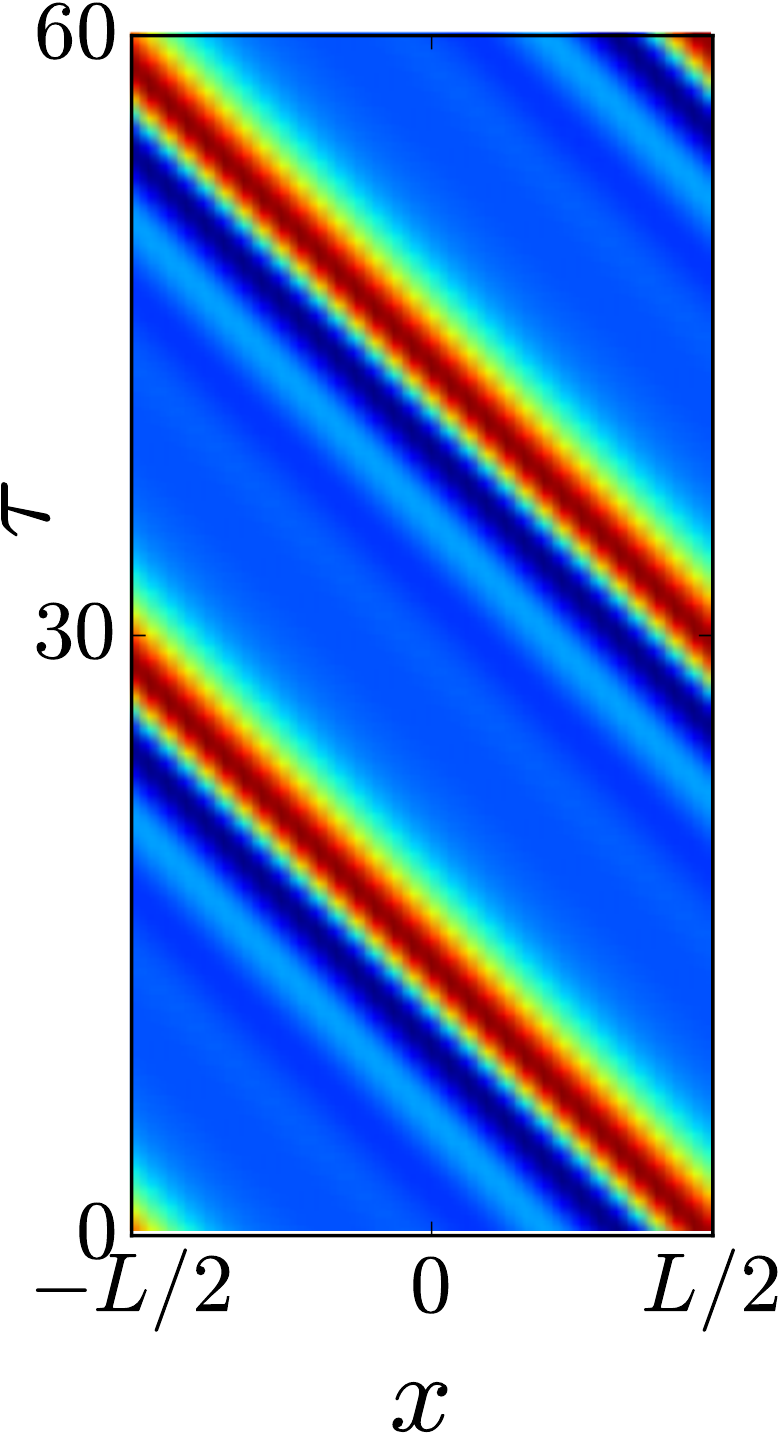}
        \put (-0.75,-1) {(b)}
    \end{overpic} \quad
    \begin{overpic}[height=0.30\textwidth]{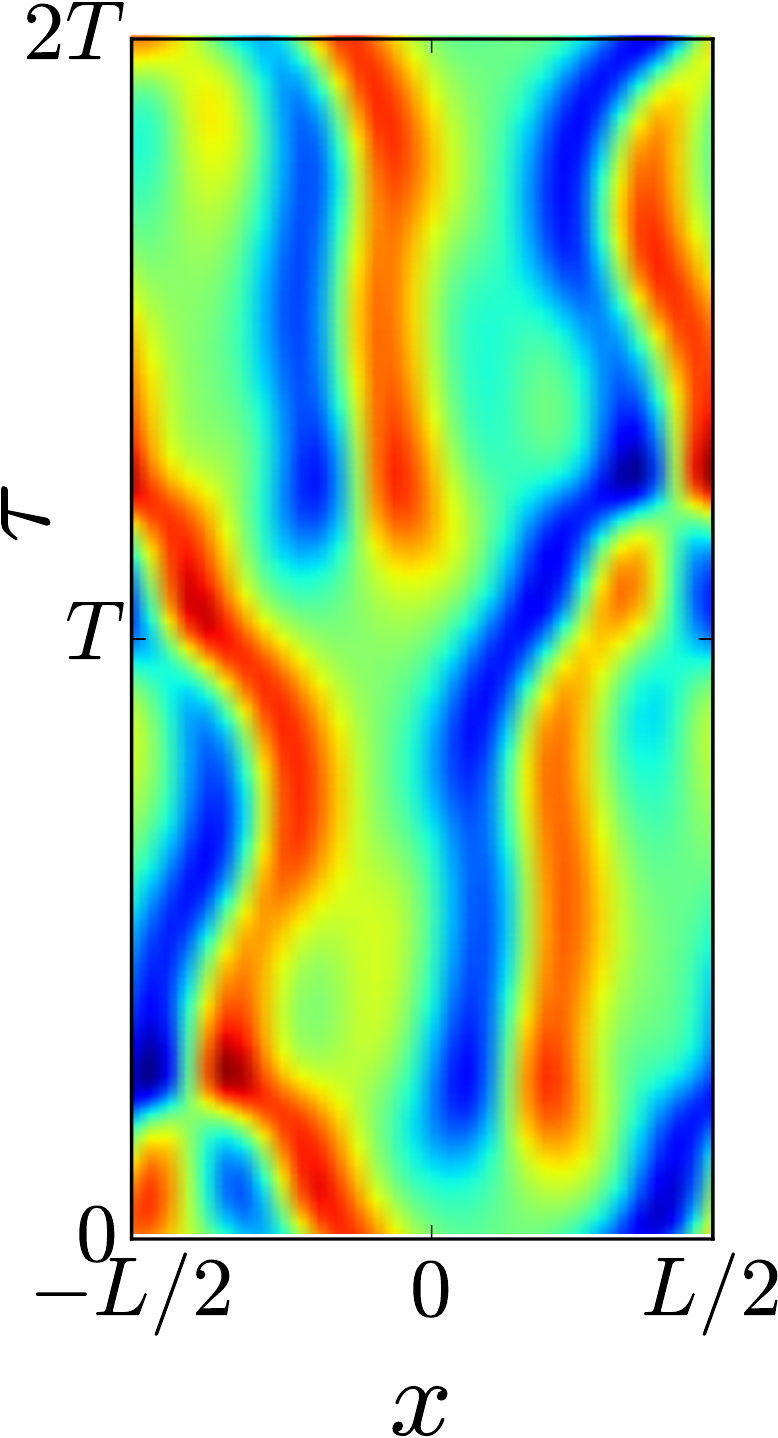}
        \put (-0.75,-1) {(c)}
    \end{overpic} \quad
    \begin{overpic}[height=0.30\textwidth]{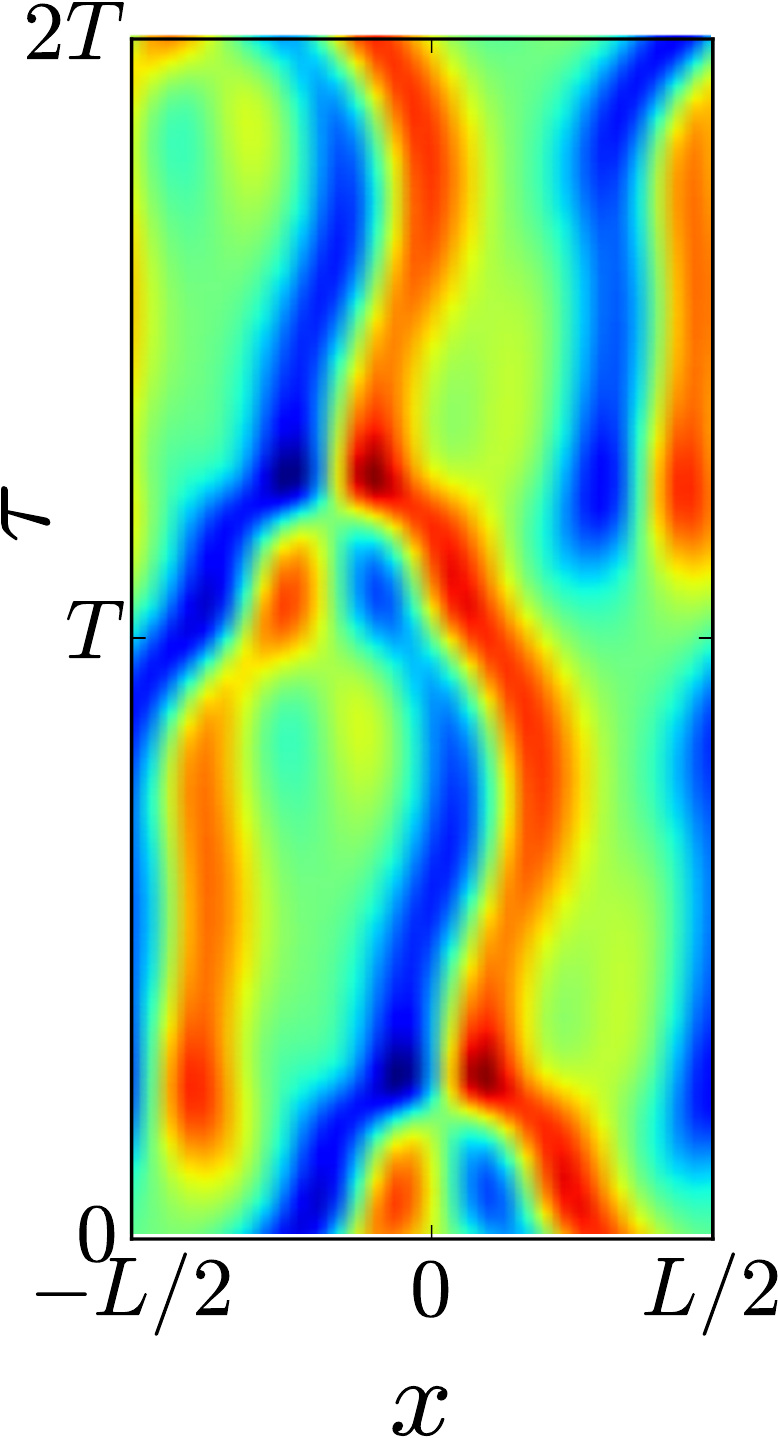}
        \put (-0.75,-1) {(d)}
    \end{overpic} \quad
    \begin{overpic}[height=0.30\textwidth]{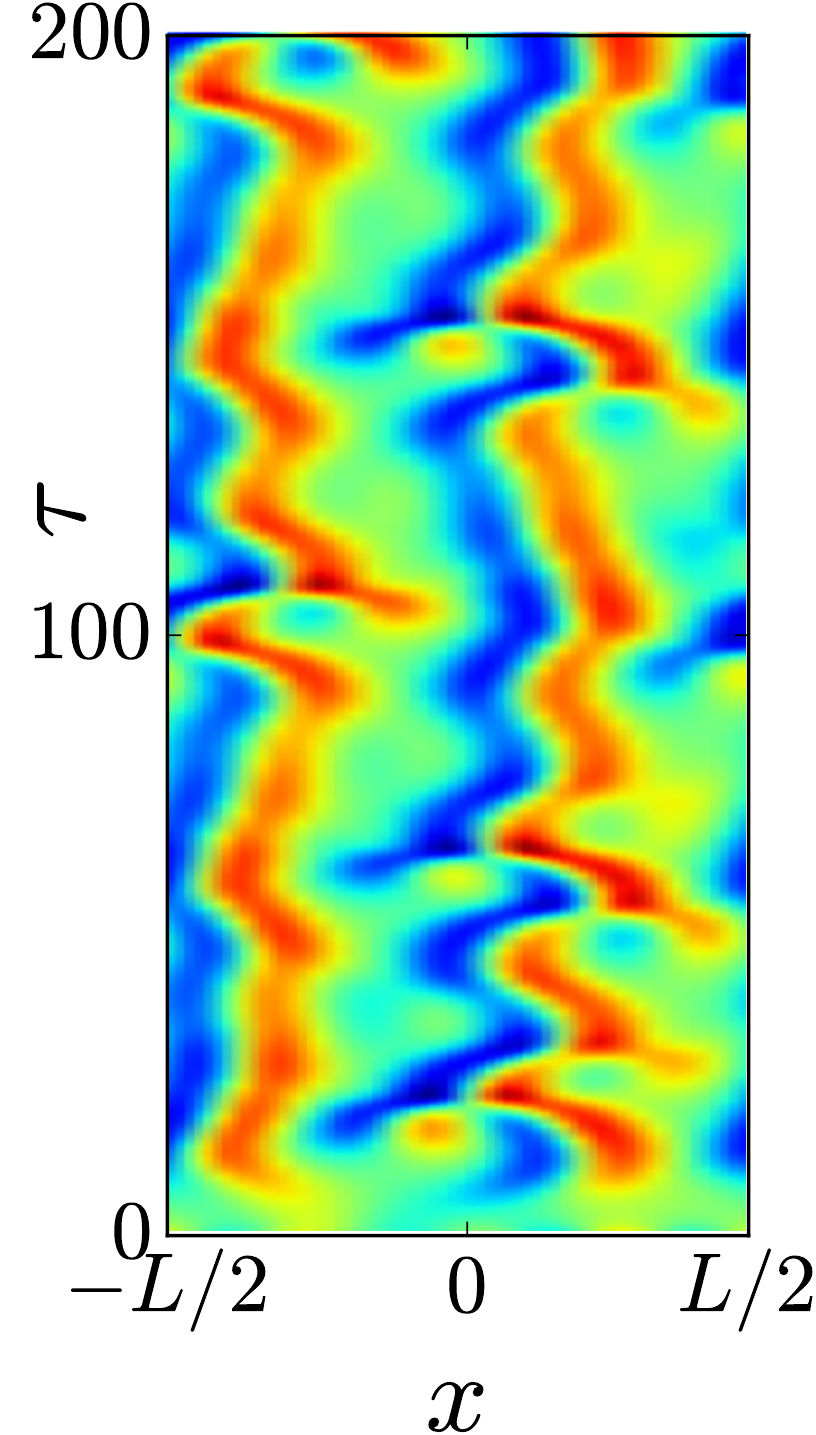}
        \put (-0.75,-1) {(e)}
    \end{overpic}
    \caption{
        Examples of invariant solutions of the \KS\ system and the
        chaotic
        flow visualized as the color coded amplitude of the scalar field
        $u(\conf, \zeit)$:
        (a) \Eqv\ $E_1$,
        (b) \Reqv\ $TW_1$,
        (c) Pre-\po\ with period $T=32.4$,
        (d) \Rpo\ with period $T=33.5$ .
        (e) Chaotic flow.
        Horizontal and vertical axes correspond to space
        and time respectively.
        System size $L=22$. The invariant solutions and their labels are
        taken from \refref{SCD07}.
    }
    \label{f-ksconf}
\end{figure}

We study the \KSe\ in one space dimension
\beq
u_\zeit = -u\,u_\conf
-u_{\conf \conf}-u_{\conf \conf \conf \conf} \,,
\label{e-ks}
\eeq
with periodic boundary condition $u(\conf,\zeit)=u(\conf+L, \zeit)$. The
real field $u(\conf, \zeit)$ is the ``flame front'' velocity\rf{siv}. The
domain size $L$ is the bifurcation parameter for the system, which
exhibits spatiotemporal chaos for sufficiently large $L$: see
\reffig{f-ksconf}\,(e) for a typical spatiotemporally chaotic trajectory
of the system at $L=22$.

We discretize the \KS\ system by Fourier expanding the field
\(
u(\conf, \zeit) = \sum_k \Fu_k (\zeit) e^{i q_k \conf}
\,,
\)
and expressing \refeq{e-ks} in terms of Fourier modes as an infinite set
of ordinary differential equations (ODEs)
\beq
\dot{\Fu}_k = ( q_k^2 - q_k^4 )\, \Fu_k
- i \frac{q_k}{2} \!\sum_{m=-\infty}^{+\infty} \!\!\Fu_m \Fu_{k-m}
\,,\quad
q_k = \frac{2 \pi k}{L}
\,.
\label{e-Fks}
\eeq

\KSe\ is \emph{Galilean invariant}: if $u(\conf,\zeit)$ is a solution, then
$v+u(\conf-v\zeit,\zeit)$, with $v$ an arbitrary constant velocity, is also a
solution. In the Fourier representation \refeq{e-Fks}, the {Galilean
invariance} implies that the zeroth Fourier mode $\Fu_0$ is decoupled
from the rest and time-invariant. Hence, we exclude $\Fu_0$ from the
\statesp\ and represent  a \KS\ state $u=u(\conf,\zeit)$  by the Fourier
series truncated at $k\!=\!N$, \ie, a $2N$\dmn\ real valued \statesp\
vector
\beq
    \ssp = (b_1, c_1, b_2, c_2, \dots, b_N, c_N)\,,
\label{e-Statesp}
\eeq
where $b_k = \Re[\Fu_k]$, $c_k = \Im [\Fu_k]$. One can rewrite
\refeq{e-Fks} in terms of this real valued \statesp\ vector, and express
the truncated set of equations compactly as
\beq
    \dot{\ssp} = \vel (\ssp )
\,.
\label{e-ODE}
\eeq
In our numerical work we use a pseudo-spectral formulation of
\refeq{e-ODE}, described here in \refappe{s-Stability}, and in
detail in the appendix of \refref{SCD07}.

Spatial translations $u(\conf, \zeit) \rightarrow u(\conf + \delta \conf,
\zeit)$ on a periodic domain correspond to $\SOn{2}$ rotations
\(
\ssp \to \matrixRep(\LieEl (\theta))\,\ssp
\)
in the \KS\ \statesp, with the
matrix representation
\beq
\matrixRep(\LieEl(\theta)) = \mathrm{diag}\left[\,{}
R(\theta),\, R(2 \theta),\, \ldots,\,
R (N \theta)\,\right]\,,
\label{e-DSO2}
\eeq
where
$\theta = 2 \pi \delta \conf / L$
and
\[
R(k \theta) =  \begin{pmatrix}
    \cos k\theta & -\sin k\theta   \\
    \sin k\theta  & ~\cos k\theta
\end{pmatrix}
\]
are $[2\!\times\!2]$ rotation matrices. \KS\ dynamics commutes with
the action of \refeq{e-DSO2}, as can be verified by checking that
\refeq{e-ODE} satisfies the equivariance relation
\beq
\vel (\ssp) = \matrixRep^{-1} (\LieEl(\theta))
			  \vel(\matrixRep(\LieEl(\theta)) \ssp)
\,.
\ee{eqvRelat}
By the translation symmetry of the \KS\ system, each solution of
PDE \refeq{e-ks}
has infinitely many dynamically equivalent copies that can be obtained by
translations \refeq{e-DSO2}. Systems with continuous symmetries thus tend
to have higher-dimensional invariant solutions: \reqva\ (traveling waves)
and \rpo s. A \emph{\reqv} evolves only along the continuous symmetry
	direction
\[
\sspTW (\zeit) = \matrixRep(\LieEl( \zeit\,\dot{\theta}_{tw}) ) \,
\sspTW (0) \,,
\]
where $\dot{\theta}_{tw}$ is a constant \phaseVel, and the suffix
$\scriptsize{tw}$ indicates that the solution is a ``traveling wave.''
A \emph{\rpo} recurs exactly at a symmetry-shifted location after
	one period
\beq
\sspRPO (0) = \matrixRep(\LieEl(- \theta_{rp} ) )
\, \sspRPO (\period{rp}) \,.
\label{e-RPO}
\eeq
\refFig{f-ksconf}\,(b) and (d) show space-time visualizations
of a \KS\ \reqv\ and a \rpo. The sole
dynamics of a \reqv\ is a constant drift along the continuous symmetry
direction, while a \rpo\ shifts by amount $\theta_{rp}$ for each repeat of
its period, and traces out a torus in the full \statesp.

The \KSe\ \refeq{e-ks} has no preferred direction, and is thus also
equivariant under the \emph{reflection} symmetry $u(\conf, \zeit)
\rightarrow - u (- \conf, \zeit)$: for each solution drifting left, there
is a reflection-equivalent solution which drifts right. In terms of
Fourier components, the reflection $\sigma$ acts as complex conjugation
followed by a negation,
whose action on vectors in \statesp\ \refeq{e-Statesp} is represented by
the diagonal matrix
\beq
\matrixRep(\sigma) = \mathrm{diag}\left[\,-1,\, 1,\,-1,\, 1,\,
\ldots, \,-1,\, 1 \right]
\,,
\label{e-DR}
\eeq
which flips signs of the real components $b_i$.
Due to this reflection symmetry, the \KS\ system can also have
strictly non-drifting \eqva\ and (pre-)\-\po s. An \emph{\eqv} is a
stationary solution
\(
\sspEQ (\zeit) = \sspEQ (0)
\,.
\)
A \emph{\po} $p$ is periodic with period \period{p},
\(
\ssp_p(0) = \ssp (\period{p})
\,,
\)
and a \emph{pre-\po} is a \rpo\
\beq
\sspPPO (0) = \matrixRep(\sigma)\,\sspPPO (\period{pp})
\label{e-PPO}
\eeq
which closes in the full \statesp\ after the second repeat,
hence we refer to it here as `pre-periodic'.

In \reffig{f-ksconf}\,(a) we show \eqv\ $E_1$ of \KSe\ (so labelled in
\refref{SCD07}).
If we were to take the mirror image of \reffig{f-ksconf}\,(a) with
respect to $\conf = 0$ line, and then interchange red and blue colors,
we would obtain the same solution; all \eqva\ belong to
the flow-invariant subspace of solutions invariant under the
reflection symmetry of the \KSe.
		Similar to equilibria, time-periodic solutions of the \KSe\ that
	    are not repeats of pre-periodic ones \refeq{e-PPO} also
	    belong to the reflection-invariant subspace. See
		\rf{Christiansen97,RCMR04,ReCi05,lanCvit07,ReChMi07}
		for examples of such solutions.
\refFig{f-ksconf}\,(b) shows a pre-periodic solution of the \KS\
system: dynamics of the second period can be obtained
from the first one by reflecting it. Both \eqva\ and pre-\po s have
infinitely many copies that can be obtained by
continuous translations, symmetric across the shifted symmetry line,
$\LieEl(\theta) \sigma \LieEl(-\theta)$.
Note that reflection $\sigma$ and translations
$\LieEl(\theta)$ do not commute:
\(
\sigma\,\LieEl(\theta) = -\LieEl(\theta)\,\sigma
\,,
\)
or, in terms of the generator of translations, the reflection reverses the
direction of the translation,
\(
\sigma\,\Lg = -\Lg\,\sigma
\).
Let $\flow{\zeit}{\ssp}$ denote the finite-time flow induced by
\refeq{e-ODE}, and let $\sspPPO$ belong to a pre-\po\  defined by
\refeq{e-PPO}. Then the shifted point
$\sspPPO' = \matrixRep(\LieEl(\theta))\,\sspPPO$ satisfies
\[
    \flow{\period{p}}{\sspPPO'} = \matrixRep(\LieEl(\theta))
                                  \matrixRep(\sigma)
                                  \matrixRep(\LieEl(-\theta))
                                  \,\sspPPO'
\,.
\]
In contrast, a \rpo\ \refeq{e-RPO} has a distinct reflected copy
$\sspRPO' = \matrixRep(\sigma) \sspRPO$ with  the reverse phase shift:
\[
    \sspRPO' (0) = \matrixRep(\LieEl(\theta_p ) )
    \, \sspRPO' (\period{p})
\,.
\]
In order to carry out our analysis, we must first eliminate all these
degeneracies. This we do by symmetry reduction, which we
describe next.

\section{Symmetry reduction}
\label{s-SymmRed}

A group orbit of state $\ssp$ is the set of all \statesp\ points
reached by
applying all symmetry actions to $\ssp$.
\emph{Symmetry reduction} is any coordinate transformation that maps
each group orbit to
a unique \statesp\ point \sspRefRed\ in the symmetry-reduced \statesp.
For the \On{2}\ symmetry considered here, we achieve this in two steps:
We first reduce continuous translation symmetry of the system by
\mslices, and then reduce the remaining reflection symmetry by
constructing an invariant polynomial basis.

To the best of our knowledge, Cartan\rf{CartanMF} was first to use
\mslices\ in purely differential geometry context and early
appearances of slicing methods in dynamical systems literature
are works of Field\rf{Field80} and Krupa\rf{Krupa90}.
Our implementation of the \mslices\ for \SOn{2}
symmetry reduction follows \refref{BudCvi14}. For a more exhaustive
review of the literature we refer the reader to \refref{DasBuch}.

Invariant polynomial or `integrity' bases\rf{gatermannHab,ChossLaut00}
are a standard tool\rf{Hilbert93,Noether15} for orbit space reduction.
They work very well in low
dimensions\rf{ChossLaut00,GL-Gil07b,SiCvi10,BuBoCvSi14}, but in high
dimensions integrity bases are high-order polynomials of the original
\statesp\ coordinates, accompanied by large numbers of nonlinear
{syzygies} that confine the
symmetry-reduced dynamics to lower-dimensional manifolds. These make the
geometry of the reduced \statesp\ complicated and hard to work with for
applications we have in mind here, such as visualizations of unstable
manifolds of invariant solutions. Even with the use of computer
algebra\rf{gatermannHab}, constructing an \On{2}-invariant integrity
basis becomes impractical for systems of dimension higher than $\sim 12$.
In spatio-temporal and fluid
dynamics applications the corresponding $n$ (Fourier series truncation)
is easily of order 10-100. The existing methods for construction of
such integrity bases are neither feasible for higher\dmn\ \statesp
s\rf{SiminosThesis} (we need to reduce symmetry for $10^5$-$10^6$\dmn\
systems\rf{GHCW07,ACHKW11}), nor helpful for reduced \statesp\
visualizations ($m$-th Fourier coefficient is usually replaced by a
polynomial of order $m$).

Here we avoid constructing such high-order \On{2} polynomial integrity
bases by a hybrid approach. We reduce the continuous symmetry by the
\fFslice\ in \refsect{s-SymmRedCont}, and \emph{then} reduce the remaining
reflection symmetry by a transformation to invariant polynomials in
\refsect{s-SymmRedDiscr}. The resulting polynomials are only
second order in the original \statesp\ coordinates, with no syzygies.

\subsection{{\SOn{2}} symmetry reduction}
\label{s-SymmRedCont}

\begin{figure}[h]
	\floatbox[{\capbeside\thisfloatsetup{capbesideposition={left,center},capbesidewidth=0.4\textwidth}}]{figure}[\FBwidth]
	{\caption{    A sketch of the full \statesp\ trajectory $\ssp(\zeit)$ (blue
			and red) projected onto the first Fourier mode subspace $(b_1, c_1)$,
			with rotation phases $\phi(\zeit_{1})$, $\phi(\zeit_{2})$
			at times  $\zeit_1$ and $\zeit_2$, see \refeq{e-fFslice}.
				In this 2\dmn\ projection we are looking at the symmetry-reduced
				\statesp\ ``from the top''; the symmetry-reduced orbit is confined to
				the horizontal half-axis $(\hat{b}_1 > 0,\, \hat{c}_1=0)$ , and the
				remaining $2N\!-\!2$ coordinates are all projected onto the origin.
			}\label{f-fFslice}}
	{
    \setlength{\unitlength}{0.5\textwidth}
    \begin{picture}(1,0.75)%
    \put(0,0){\includegraphics[width=\unitlength]{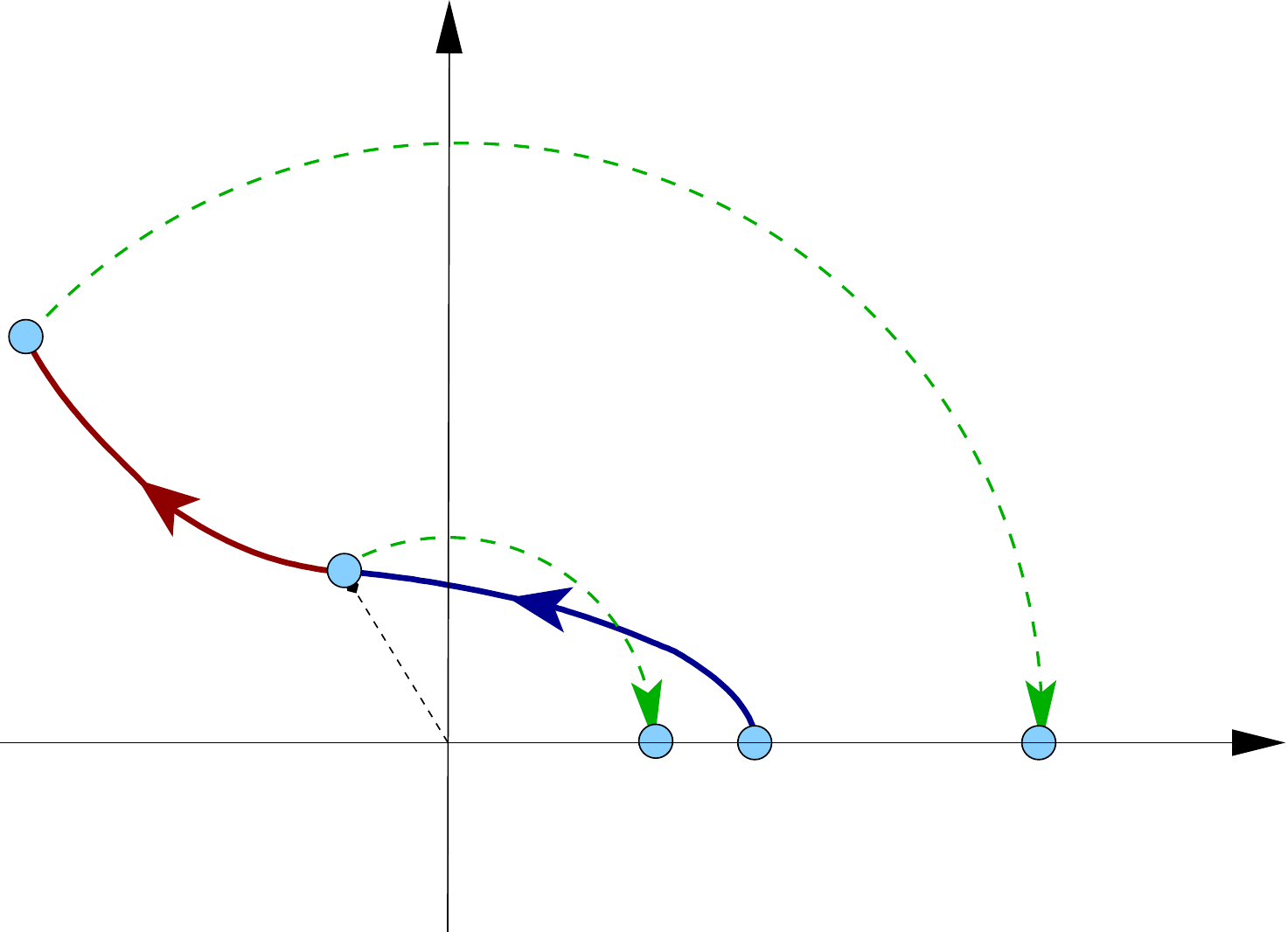}}%
    \put(0.001,0.5){\color[rgb]{0,0,0}\makebox(0,0)[lb]{\smash{$\ssp(\zeit_2)$}}}%
    \put(0.18,0.32){\color[rgb]{0,0,0}\makebox(0,0)[lb]{\smash{$\ssp(\zeit_1)$}}}%
    \put(0.37,0.32){\color[rgb]{0,0,0}\makebox(0,0)[lb]{\smash{$\phi (\zeit_1)$}}}%
    \put(0.95,0.08){\color[rgb]{0,0,0}\makebox(0,0)[lb]{\smash{$b_1$}}}%
    \put(0.42,0.07){\color[rgb]{0,0,0}\makebox(0,0)[lb]{\smash{$\sspRed(\zeit_1)$}}}%
    \put(0.77,0.07){\color[rgb]{0,0,0}\makebox(0,0)[lb]{\smash{$\sspRed(\zeit_2)$}}}%
    \put(0.28,0.70){\color[rgb]{0,0,0}\makebox(0,0)[lb]{\smash{$c_1$}}}%
    \put(0.59264739,0.58){\color[rgb]{0,0,0}\makebox(0,0)[lb]{\smash{$\phi (\zeit_2)$}}}%
    \put(0.58,0.07){\color[rgb]{0,0,0}\makebox(0,0)[lb]{\smash{$\ssp(\zeit_0)$}}}%
    \end{picture}
		}
	\end{figure}

Following \refref{BudCvi14}, we reduce the \SOn{2} symmetry of the
\KSe\ by implementing the \emph{\fFslice} method, \ie, by rotating the Fourier
modes as
\beq
    \sspRed(\zeit) = \matrixRep(\LieEl(\phi(\zeit))^{-1})\,\ssp(\zeit)
\,,
\label{e-fFslice}
\eeq
where $\phi (\zeit) = \arg (\Fu_1 (\zeit))$
is the phase of the first Fourier mode.
This transformation exists
as long as the first mode in the Fourier expansion \refeq{e-fFslice} does
not vanish, $b_1^2+c_1^2 > 0$,
and its effect is to fix the phase of the first Fourier mode to zero for
all times, as illustrated in \reffig{f-fFslice}. The
\SOn{2}-reduced \statesp\ is one dimension lower than the full
\statesp, with coordinates
\beq
\sspRed = (\hat{b}_1, 0, \hat{b}_2, \hat{c}_2,
           \ldots \hat{b}_N, \hat{c}_N)\,.
\label{e-sspRed}
\eeq
The dynamics within the \fFslice\ is given by
\beq
    \dot{\sspRed} = \velRed (\sspRed)
                  = \vel(\sspRed) - \frac{\dot{c}_1}{\hat{b}_1} \,\Lg \sspRed
\,,
\label{e-velRed}
\eeq
where $\Lg$ is the generator of infinitesimal \SOn{2} transformations,
$\matrixRep(\LieEl(\theta)) = \exp \Lg \theta$, and $\dot{c}_1$ is full
\statesp\ orbit's out-of-slice velocity, the second element of the
velocity field \refeq{e-ODE}.
		Symmetry-reduced \statesp\ velocity \refeq{e-velRed} diverges
	    when the amplitude $\hat{b}_1$ of the first Fourier mode tends
	    to $0$. If $\hat{b}_1$ were $0$, then the transformation
	    \refeq{e-fFslice} would no longer be uniquely defined. However,
	    our experience had been such that this does not happen for
	    generic trajectories of a chaotic system; and the singularity
	    in the vicinity of $\hat{b}_1 = 0$ can be regularized by a
	    time-rescaling transformation\rf{BudCvi14}.
For further details we refer the reader to
\refrefs{DasBuch,BudCvi14,BuBoCvSi14}.

\subsection{{\On{2}} symmetry reduction}
\label{s-SymmRedDiscr}

Our next challenge is to devise a transformation from \refeq{e-sspRed} to
discrete-symmetry-reduced coordinates, where the equivariance under
reflection is also reduced.
Consider the action of reflection on the \SOn{2}-reduced \statesp. In
general, a slice is an arbitrarily oriented hyperplane, and action of the
reflection $\sigma$ can be rather complicated: it maps points within the
{\slicePlane} into points outside of it, which then have to be rotated
into the slice.
However, the action of $\sigma$ on the \fFslice\ is particularly simple.
Reflection operation $\matrixRep(\sigma)$ of \refeq{e-DR} flips the sign of the
first \SOn{2}-reduced \statesp\ coordinate in \refeq{e-sspRed}, \ie, makes
the phase of the first Fourier mode $\pi$. Rotating back into the slice
by \refeq{e-fFslice}, we find that within the \fFslice, the reflection
acts by alternating the signs of even (real part) and odd (imaginary
part) Fourier modes:
\bea
    \matrixRepRed (\sigma) &=& \matrixRep(\LieEl(- \pi))
                             \matrixRep(\sigma) \continue
                           &=& \mathrm{diag}\left[\,1,\,-1,\,-1,\, 1,
                                            \,1,\,-1,\,-1,\, 1,\, 1,\,
                                          \ldots \right]
\,.
\label{e-DRRed}
\eea
The action on the slice coordinates
(where we for brevity omit all terms whose signs do not
change under reflection) is thus
\bea
&\matrixRepRed (\sigma)& (\hat{b}_2, \hat{c}_3, \hat{b}_4,
\hat{c}_5, \hat{b}_6, \hat{c}_7, \ldots )
\continue
&& = (-\hat{b}_2, -\hat{c}_3, -\hat{b}_4,
-\hat{c}_5, -\hat{b}_6, -\hat{c}_7, \ldots )
\,.
\label{e-EvenOdd}
\eea

Our task is now to construct a transformation to a set of
coordinates invariant under (\ref{e-EvenOdd}).
One could declare a half of the symmetry-reduced \statesp\ to be a
`fundamental domain'\rf{DasBuch}, with segments of orbits that exit it
brought back by reflection, but this makes orbits appear discontinuous
and the dynamics hard to visualize. Instead, here we shall reduce the
reflection symmetry by constructing polynomial invariants of coordinates
(\ref{e-EvenOdd}). Squaring (or taking absolute value of) each
sign-flipping coordinate in (\ref{e-EvenOdd}) is not an option, since
such coordinates would be invariant under every individual sign change of
these coordinates, and that is not a symmetry of the system. We are
allowed to impose \emph{only one} condition to reduce the 2-element group
orbit of the discrete reflection subgroup of \On{2}.
How that can be achieved is suggested by Miranda and
Stone\rf{GL-Mir93,GL-Gil07b} reduction of $\Ztwo$ symmetry
\(
(x, y, z) \rightarrow (-x, -y, z)
\)
of the Lorenz flow. They construct the
symmetry-reduced ``proto-Lorenz system'' by transforming coordinates
to the polynomial basis
\beq
    u = x^2 - y^2 \,,\quad v = 2xy\,, \quad z = z
\,.
    \label{e-InvPolLorenz}
\eeq
The $x$ coordinate can be recovered from $u$ and $v$ of
\refeq{e-InvPolLorenz} up to a choice of sign, \ie, up to the original
reflection symmetry. We extend this approach in order to achieve a
$2\!-\!\mbox{to}\!-\!1$ symmetry reduction for \KS\ system:
we construct the first coordinate from
squares, but then `twine' the successive sign-flipping terms
\(
(\hat{b}_2, \hat{c}_3, \hat{b}_4,
\hat{c}_5, \hat{b}_6, \hat{c}_7, \ldots )
\)
into second-order invariant polynomials basis set
\bea
    && (p_2, p_3, p_4, p_5, p_6, \ldots) \nnu\\
    && = (\hat{b}_2^2 - \hat{c}_3^2, \,
         \hat{b}_2 \hat{c}_3, \,
         \hat{b}_4 \hat{c}_3, \,
         \hat{b}_4 \hat{c}_5, \,
         \hat{b}_5 \hat{c}_6, \,
         \ldots)
\,.
\label{e-RefInvs}
\eea
The original coordinates can be recovered
recursively by the $1\!-\!\mbox{to}\!-\!2$  inverse transformation
\bea
    b_2 &=& \pm \sqrt{\frac{p_2 + \sqrt{p_2^2 + 4 p_3^2}}{2}} \continue
    c_3 &=& p_3 / b_2\,,\quad
    b_4 \,=\, p_4 / c_3\,,\quad
    c_5 \,=\, p_5 / b_4\,,\quad\cdots
\,.
\nonumber
\eea

To summarize: we first reduce the group orbits generated by the continuous
\SOn{2}\ symmetry subgroup by implementing the {\fFslice} \refeq{e-fFslice},
and then reduce the group orbits of the discrete 2-element reflection subgroup
by replacing the sign-changing coordinates (\ref{e-EvenOdd}) with the
invariant polynomials (\ref{e-RefInvs}).
The final \On{2} symmetry-reduced coordinates are
\beq
    \sspRefRed = 
    (\hat{b}_1, 0,
    \hat{b}_2^2 - \hat{c}_3^2,
    \hat{c}_2,
    \hat{b}_3,
    \hat{b}_2 \hat{c}_3,
    \hat{b}_4 \hat{c}_3,
    \hat{c}_4,
    \hat{b}_5,
    \ldots )\,.
\label{e-sspRefRed}
\eeq
Here pairs of orbits related by reflection $\sigma$ are mapped into a
single orbit, and $\hat{c}_1$ is identically
set to $0$ by continuous symmetry reduction, thus the symmetry-reduced
\statesp\ has one dimension less than the full \statesp.

The symmetry-reduced \statesp\ \refeq{e-sspRefRed} retains all
physical information of the \KS\ system:
\reqva\ and \rpo s of the original system
become \eqva\ and \po s in the symmetry-reduced \statesp\
\refeq{e-sspRefRed}, and pre-\po s close after one period.
For this reason, in what follows we shall refer to both
\rpo s and pre-\po s as `\po s', unless we comment on their
specific symmetry properties.

\section{Unstable manifolds of \po s}
\label{s-UnstMan}

In order to demonstrate the utility, and indeed, the necessity of the
\On{2}\ symmetry reduction, we now investigate the transition to chaos in
the neighborhood of a short \KS\ {\ppo}, focusing on the parameter range
$L \in [21.0, 21.7]$.
Our method yields a symmetry-reduced velocity field
$\velRefRed (\sspRefRed) = \dot{\sspRefRed}$
and a finite-time flow
$\flowRefRed{\zeit}{\sspRefRed(0)} = \sspRefRed(\zeit)$
in the symmetry-reduced \statesp\ \refeq{e-sspRefRed}.
Although we can obtain $\velRefRed (\sspRefRed)$ by chain rule, we find
its numerical integration unstable, hence in practice we obtain
$\velRefRed (\sspRefRed)$ and $\flowRefRed{\zeit}{\sspRefRed}$ from the
\fFslice\ by applying the appropriate \jacobianMs, as described in
\refappe{s-Stability}.

\begin{table}
	\caption{\label{t-EeigVals}
		The two leading non-marginal Floquet multipliers
		$\ExpaEig = \exp(\period{} \mu + i\theta)$ of \po s \primeOrb{0},
		\primeOrb{1}, \primeOrb{2} for system sizes $L$ studied here. Dash --
		indicates that the orbit is not found for the corresponding system size.
	}
	\begin{center}
		\begin{tabular}{c || c c | c c | c c}
			& \multicolumn{2}{c}{\primeOrb{0}}
			& \multicolumn{2}{c}{\primeOrb{1}}	
			& \multicolumn{2}{c}{\primeOrb{2}} \\
			L & $\mu$ &$\theta$
			& $\mu$ &$\theta$
			& $\mu$ &$\theta$ \\
			\hline
			21.25 & $6.443 \times 10^{-4}$ & $\pm 2.177$
			& \multicolumn{2}{c |}{--}
			& \multicolumn{2}{c}{--} \\
			21.30 & $1.839 \times 10^{-3}$ & $\pm 2.158$
			& \multicolumn{2}{c |}{--}
			& \multicolumn{2}{c}{--} \\
			21.36 & $1.839 \times 10^{-3}$ & $\pm 2.158$
			& $5.854 \times 10^{-3}$ & $0$
			& $-1.623 \times 10^{-3}$ & $\pm 0.3098$ \\
			& &
			& $- 8.357 \times 10^{-3}$ & $0$
			& & \\
			21.48 & $7.638 \times 10^{-3}$ & $\pm 2.097$
			& $1.307 \times 10^{-2}$ & $ 0 $
			& \multicolumn{2}{c}{--} \\
			& &
			& $- 1.234 \times 10^{-2}$ & $0$
			& & \\
			21.70 & $1.739 \times 10^{-2}$ & $\pm 2.044$
			& $2.521 \times 10^{-2}$ & $0$
			& \multicolumn{2}{c}{--} \\
			& &
			& $4.157 \times 10^{-3}$ & $\pi$
			& &
		\end{tabular}
	\end{center}
\end{table}

At $L=21.0$, the \KS\ system has a stable \po\ \primeOrb{0},
which satisfies
\(
    \sspRefRed_{\primeOrb{0}} = \flowRefRed{\period{\primeOrb{0}}}{
                                      \sspRefRed_{\primeOrb{0}}}
\)
for any point $\sspRefRed_{\primeOrb{0}}$ on the \po\ $\primeOrb{0}$.
Linear stability of a \po\  is described by the Floquet
multipliers
$\ExpaEig_i$ and Floquet vectors $\VRefRed_i$,1
which are the eigenvalues and eigenvectors of the \jacobianM\
$\jMpsRefRed_{p}$
of the finite-time flow
$\flowRefRed{\period{p}}{\sspRefRed_{p}}$
\[
    \jMpsRefRed_{p} \VRefRed_i = \ExpaEig_i \VRefRed_i \,.
\]

Each \po\ has at least one marginal Floquet multiplier $\ExpaEig_v = 1$,
corresponding to the velocity field direction. When $L < 21.22$,
all other Floquet multipliers of \primeOrb{0}\ have absolute values less
than $1$. At $L \approx 21.22$, leading
complex pair of Floquet multipliers
$\ExpaEig_{1, 2}$ crosses the unit circle, and the corresponding
eigenplane
spanned by the real and imaginary parts of $\VRefRed_1$ develops
`spiral out' dynamics that connects to a $2$-torus.

\begin{figure}[h]
	\centering
	\begin{overpic}[height=0.45\textwidth]{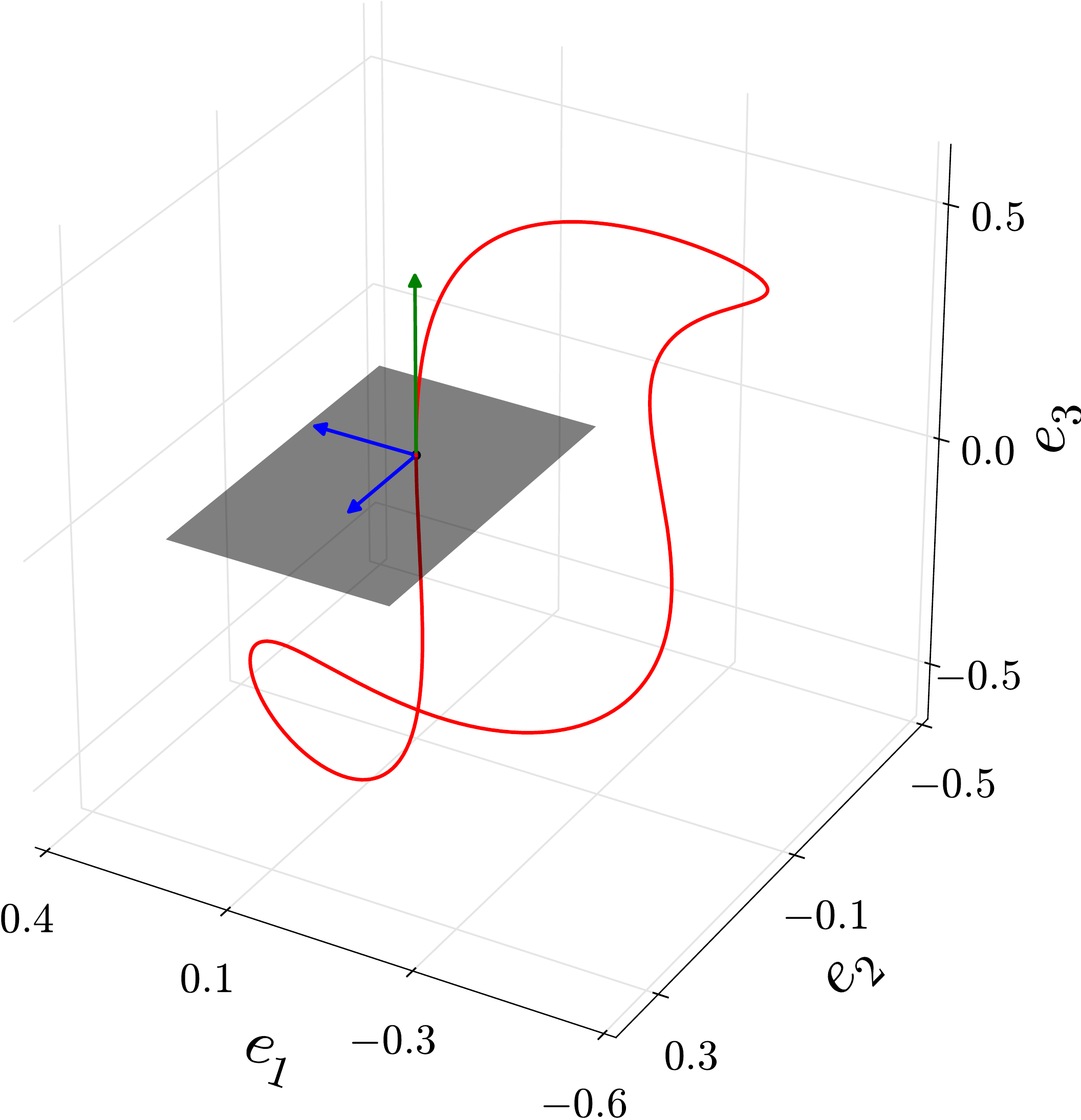}
		\put (-0.75,-1) {(a)}
	\end{overpic} \quad
	\begin{overpic}[height=0.45\textwidth]{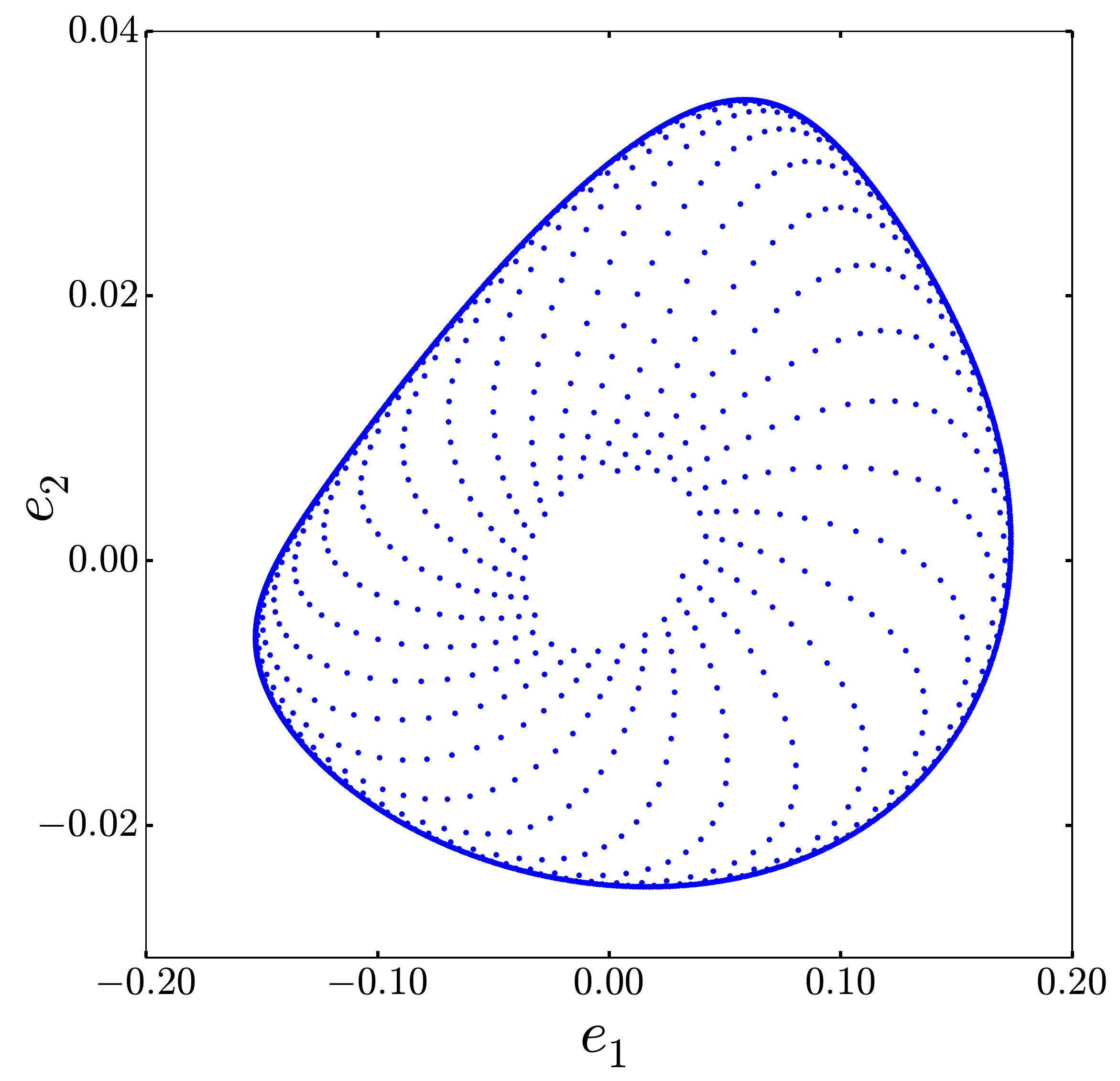}
		\put (-0.75,-1) {(b)}
	\end{overpic} \quad
	\caption{\label{f-Psect}
		(a) Pre-\po\ \primeOrb{0}\ (red), its velocity field
		$\velRefRed(\sspRefRed_{\primeOrb{0}})$ at the starting point
		(green), orthogonal vectors that span the eigenplane
		corresponding to the leading Floquet vectors (blue) and
		the {\PoincSec} hyperplane (gray, transparent).
		(b) Spiral-out dynamics of a single trajectory
		in the {\PoincSec} projected onto $(e_1, e_2)$ plane,
		system size $L=21.25$.
	}
\end{figure}

In order to study dynamics within the neighborhood of $\primeOrb{0}$, we
define a {\PoincSec} as the hyperplane of points
$\sspRefRed_{\PoincS}$
in an open
neighborhood of $\sspRefRed_{\primeOrb{0}}$, orthogonal to the tangent
$\velRefRed (\sspRefRed_{\primeOrb{0}})$ of the orbit at the {\PoincSec} point,
\beq
    (\sspRefRed_{\PoincS} - \sspRefRed_{\primeOrb{0}})
        \cdot \velRefRed (\sspRefRed_{\primeOrb{0}}) = 0
    \quad \mbox{and} \quad
    || \sspRefRed_{\PoincS} - \sspRefRed_{\primeOrb{0}} || < \alpha
    \,,
\label{e-Psect}
\eeq
where $||.||$ denotes the Euclidean (or $L2$) norm, and the threshold
$\alpha$ is empirically set to $\alpha = 0.9$ throughout.
	  The locality condition in \refeq{e-Psect} is a computationally
	  convenient way
	  to avoid Poincar\'e section border\rf{atlas12,DasBuch}, defined
	  as the set of points $\sspRefRed_{\PoincS}^*$ that satisfy the hyperplane
	  condition
$(\sspRefRed_{\PoincS}^* - \sspRefRed_{\primeOrb{0}})
  \cdot \velRefRed (\sspRefRed_{\primeOrb{0}}) = 0$	
	  , but their orbits do not intersect this hyperplane
	  transversally, \ie\
$\velRefRed (\sspRefRed_{\PoincS}^*)
 \cdot \velRefRed (\sspRefRed_{\primeOrb{0}}) = 0$.
	 
From here on, we study the discrete time dynamics induced by the
flow on the Poincar\'e section \refeq{e-Psect}, as visualized
in \reffig{f-Psect}\,(a).

In \reffig{f-Psect} and the rest of the \statesp\ projections of this
paper, projection bases are constructed as follows:
Real and imaginary parts of the Floquet vector $\VRefRed_{1}$ define
an ellipse
$\Re [\VRefRed_{1}] \cos \phi + \Im [\VRefRed_{1}] \sin \phi $
in the neighborhood of $\sspRefRed_{\primeOrb{0}}$, and we pick as the first two
projection-subspace spanning vectors the principal axes of this ellipse.
As the third vector we take the velocity field
$\velRefRed (\sspRefRed_{\primeOrb{0}})$, and the projection bases
$(e_1, e_2, e_3)$ are found by orthonormalization of these vectors
via the Gram-Schmidt procedure. All \statesp\ projections are centered
on $\sspRefRed_{\primeOrb{0}}$, \ie, $\sspRefRed_{\primeOrb{0}}$ is
the origin of all {\PoincSec} projections.

\begin{figure}[h]
	\floatbox[{\capbeside\thisfloatsetup{capbesideposition={left,center},capbesidewidth=0.3\textwidth}}]{figure}[\FBwidth]
	{\caption{Initial points (black) on the {\PoincSec} for
			unstable manifold computation and their first (red),
			second (green), and third (blue) returns. Inset: zoomed out
			view of the initial points and their first three returns.}\label{f-UnstMan21p31b}}
	{
		\begin{overpic}[width=0.6\textwidth]{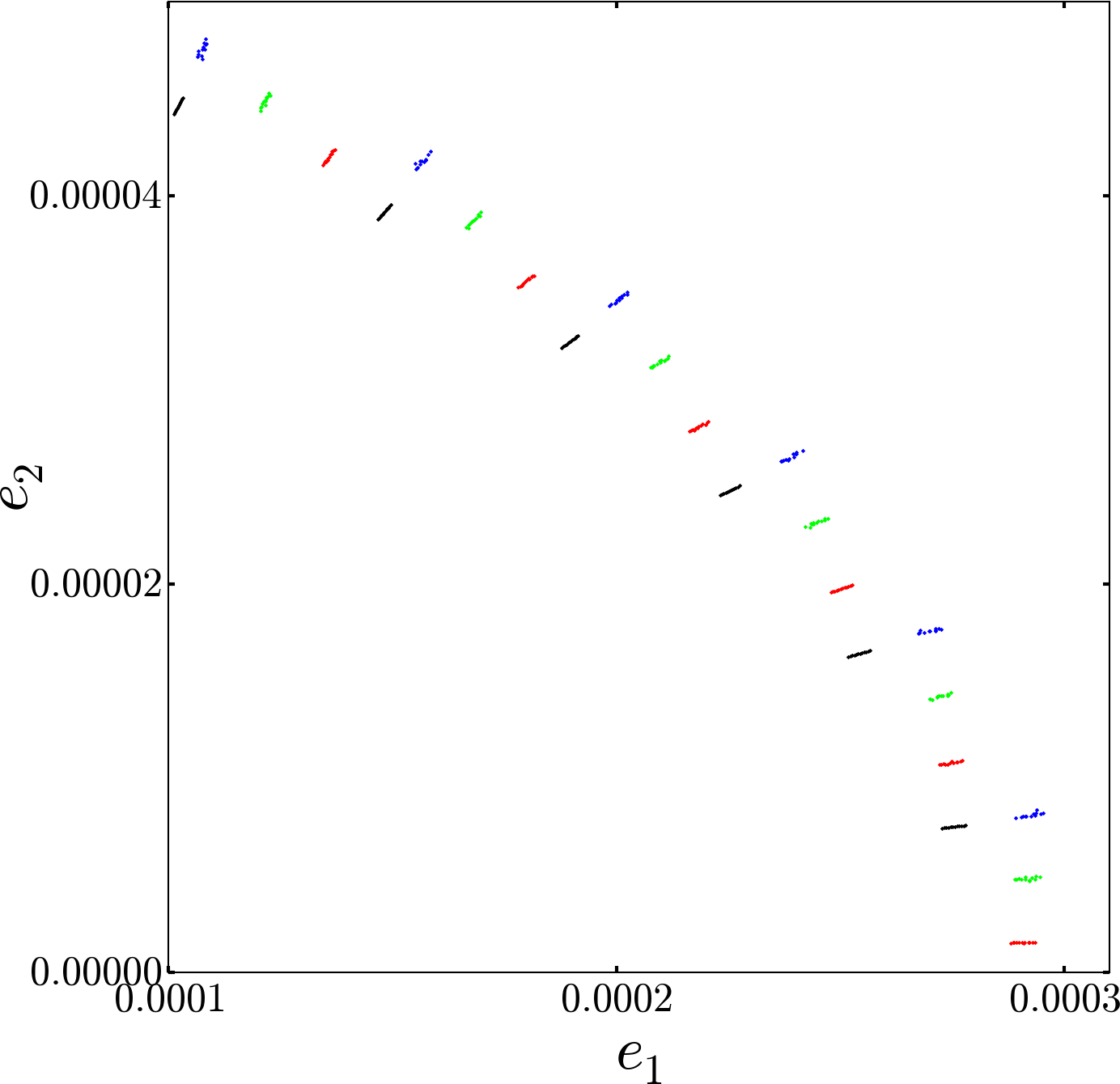}
			\put (16, 12){\includegraphics[height=0.28\textwidth]{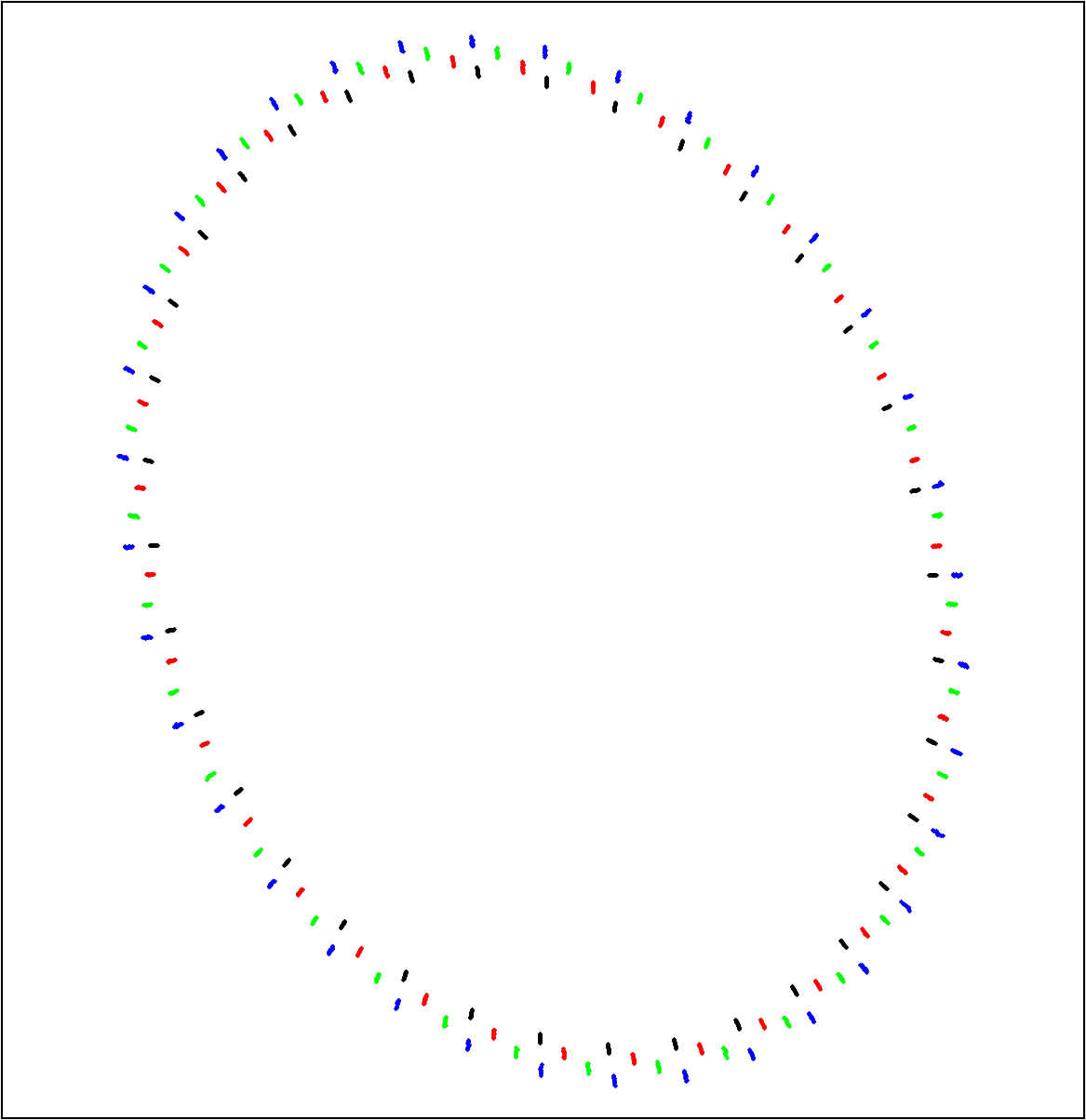}}
		\end{overpic}
	}
\end{figure}

As an example, we follow a single trajectory starting from
$\sspRefRed_{\primeOrb{0}} + 10^{-1} \Re[\VRefRed_1] $ as it connects to the
2-torus surrounding the \po\  in \reffig{f-Psect}\,(b).
For \reffig{f-Psect}\,(b) and
all figures to follow, the two leading non-marginal Floquet multipliers
of $\primeOrb{0}$, $\primeOrb{1}$ and $\primeOrb{2}$ are listed in
\reftab{t-EeigVals}.
For system size $L = 21.25$ the complex unstable
Floquet multiplier pair is nearly marginal, $|\ExpaEig_{1,2}|= 1.00636$,
hence the spiral-out is very slow.
\begin{figure}[h]
	\centering
	\includegraphics[width=0.9\textwidth]{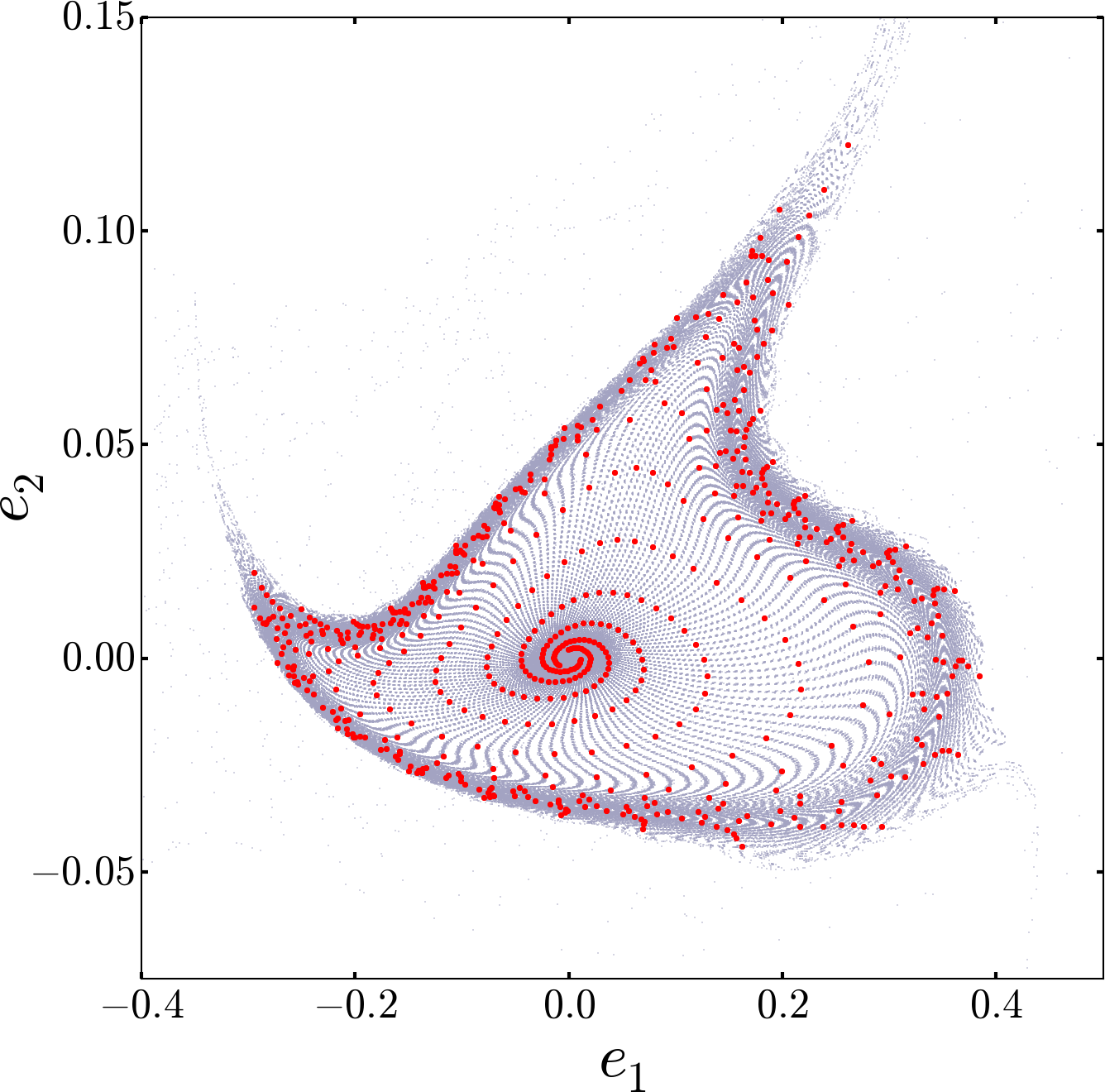}
	\caption{\label{f-UnstMan21p31a}
		Unstable manifold (gray) of $\primeOrb{0}$ on the {\PoincSec}
		\refeq{e-Psect} and an individual trajectory (red)
		within, system size $L=21.30$.
	}		
\end{figure}
Assume that $\delta \sspRefRed(0)$ is a small perturbation to
$\sspRefRed_{\primeOrb{0}}$ that lies in the plane spanned by
$(\Re[\VRefRed_1], \, \Im[\VRefRed_1])$.
Then there exists a coefficient vector $c = (c_1, c_2)^T$, with
which we can express
$\delta \sspRefRed (0)$ in this plane as
\beq
    \delta \sspRefRed (0) = W c \,,
\label{e-Perturb}
\eeq
where $W = [\Re[\VRefRed_1], \, \Im[\VRefRed_1]]$ 
has real and imaginary
parts of the Floquet vector $\VRefRed_1$ on its columns.
Without a loss of generality, we can rewrite $c$ as
$ c = \delta r R( \theta) c^{(1)} $, where
$ c^{(1)} = (1, 0)^T$ and
$R(\theta)$ is a $[2\!\times\!2]$ rotation matrix.
Thus \refeq{e-Perturb} can be expressed as
$\delta \sspRefRed (0) = \delta r W R(\theta) c^{(1)}$.
In the linear approximation, discrete time dynamics
$\delta \sspRefRed (n \period{\primeOrb{0}})$ is given by
\beq
\delta \sspRefRed (n \period{\primeOrb{0}}) = |\ExpaEig_1|^n \delta r W
                                        R(\theta - n \arg \ExpaEig_1)
                                        c^{(1)}  \,,
\label{e-DiscreteTime}
\eeq
which can then be projected onto the {\PoincSec}
\refeq{e-Psect} by acting from the left with the projection operator
\beq
    \ProjPsect (\sspRefRed_{\PoincS}) =
    \matId - \frac{\velRefRed(\sspRefRed_{\PoincS}) \otimes
                  \velRefRed(\sspRefRed_{\primeOrb{0}})}{
                  \inprod{\velRefRed(\sspRefRed_{\PoincS})}{
		                  \velRefRed(\sspRefRed_{\primeOrb{0}})}
                                } \,, \label{e-ProjPsect}
\eeq
computed at $\sspRefRed_{\PoincS} = \sspRefRed_{\primeOrb{0}}$.
In \refeq{e-ProjPsect}, 
$\otimes$ denotes the outer product.
Defining
	  $\delta \sspRefRed_{\PoincS} \equiv
	  \ProjPsect (\sspRefRed_{\PoincS}) \delta \sspRefRed$
	  for a small perturbation
	  $\delta \sspRefRed$ to the point $\sspRefRed_{\PoincS}$
	  on the Poincar\'e
	  section, discrete
time dynamics of $\delta \sspRefRed_{\PoincS}$ in
the {\PoincSec} is given by
\beq
\delta \sspRefRed_{\PoincS} [n] = |\ExpaEig_1|^n \delta r W_{\PoincS}
R(\theta - n \arg \ExpaEig_1)
c^{(1)}  \,,
\label{e-DiscreteTimePsect}
\eeq
where
$W_{\PoincS} = [\Re[\VRefRed_{1, \PoincS}], \, \Im[\VRefRed_{1, \PoincS}]]
= \ProjPsect(\sspRefRed_{\primeOrb{0}}) W$,
and $n$ is the discrete time variable counting returns to the
{\PoincSec}. In the {\PoincSec}, the solutions
\refeq{e-DiscreteTimePsect}  define ellipses which expand and rotate
respectively by factors of $|\ExpaEig_1|$ and $\arg \ExpaEig_1$ at each
return. In order to resolve the unstable manifold, we start trajectories
on an elliptic band parameterized by $(\delta,\phi)$, such that the
starting point in the band comes to the end of it on the first return,
hence totality of these points cover the unstable manifold in the linear
approximation. Such set of perturbations are given by
\beq
    \delta \sspRefRed_{\PoincS} (\delta, \phi) =
    \epsilon |\ExpaEig_1|^\delta W_{\PoincS} R(\phi) c^{(1)}
    \,, \quad \mbox{where } \delta \in [0, 1) \,, \,
                             \phi \in [0, 2 \pi )
\,,
\label{e-InitUnstMan}
\eeq
and $\epsilon$ is a small number. We set $\epsilon = 10^{-3} $ and
discretize \refeq{e-InitUnstMan}
by taking $12$ equidistant points in $[0, 1)$ for $\delta$ and
$36$ equidistant points in $[0, 2 \pi)$ for $\phi$ and integrate
each
$\sspRefRed_{\primeOrb{0}} + \delta \sspRefRed_{\PoincS} (\delta, \phi)$
forward in time. \refFig{f-UnstMan21p31a} shows the unstable
manifold of $\primeOrb{0}$ resolved by this procedure at system size
$L=21.30$, for which the torus surrounding $\primeOrb{0}$
appears to be unstable
	  as the points approaching to it first slow down and then
	  leave the neighborhood in transverse direction.
	  In order to illustrate this better, we marked
	  an individual trajectory in \reffig{f-UnstMan21p31a} color red.
In \reffig{f-UnstMan21p31b} we show initial points that go
into the calculation, and their first three returns in order
to illustrate the principle of the method.

\begin{figure}[h]
	\centering
	\includegraphics[width=0.9\textwidth]{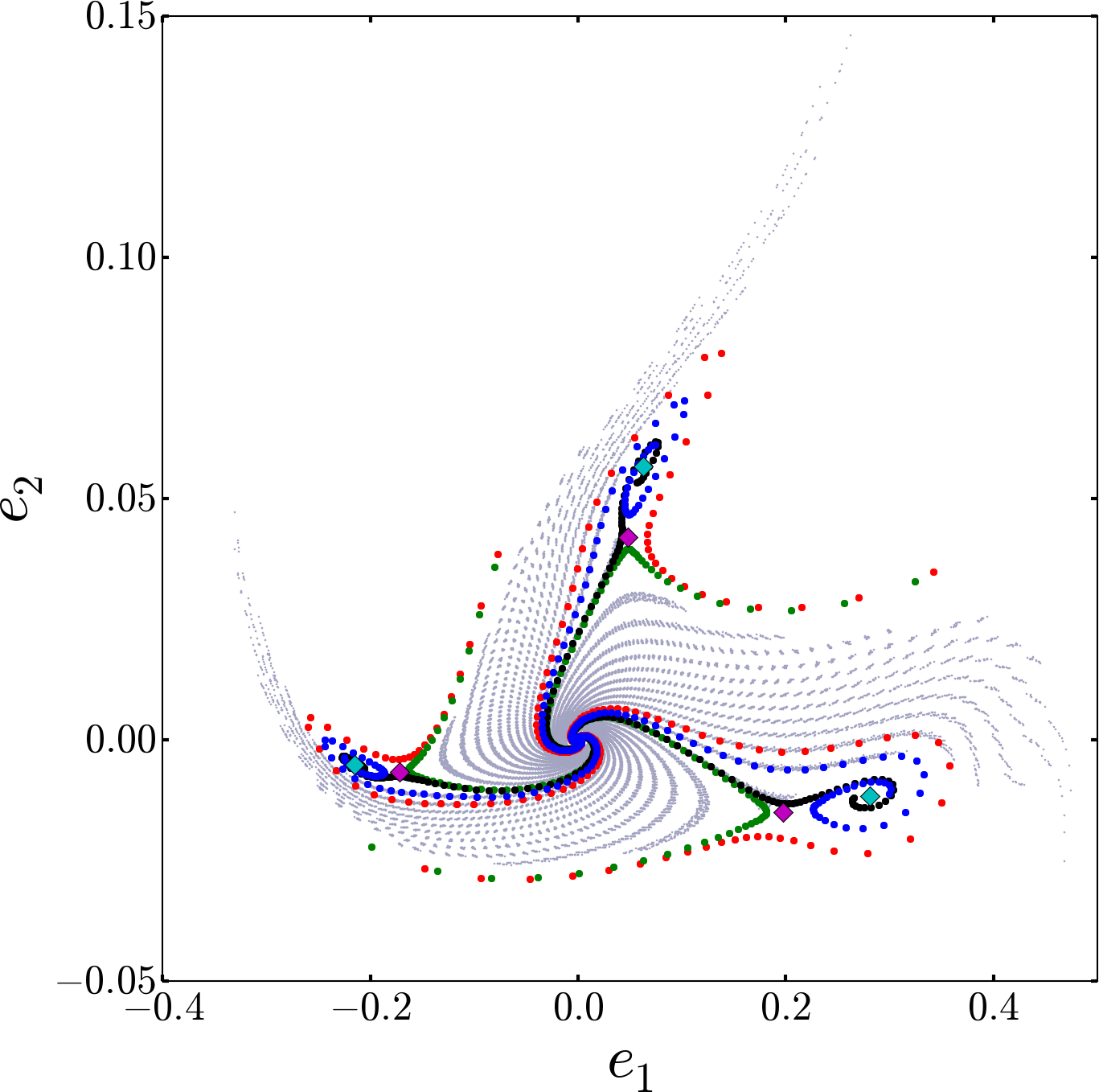}
	\caption{\label{f-Connectionsa}
		Unstable manifold (gray) of $\primeOrb{0}$ on the {\PoincSec}
		\refeq{e-Psect} at $L=21.36$. Colored dots
		correspond to different individual trajectories within
		the unstable manifold, with qualitatively different
		properties. Diamond shaped markers correspond to the
		period-3 orbits $\primeOrb{1}$ (magenta) and $\primeOrb{2}$ (cyan).
	}		
\end{figure}

As we continue increasing the system size, we find that
at $L \approx 21.36$, trace of the invariant torus disappears
and two new \po s $\primeOrb{1}$ and $\primeOrb{2}$ emerge in the
neighborhood of $\primeOrb{0}$. Both of these orbits appear as period~3
\po s in the Poincar\'e map. While
$\primeOrb{1}$ is unstable
(found by a Newton search),
$\primeOrb{2}$ is initially
stable with a finite basin of
attraction. The unstable manifold of $\primeOrb{0}$ connects
heteroclinically  to the stable manifolds of $\primeOrb{1}$ and
$\primeOrb{2}$.
As we show in
\reffig{f-Connectionsa}, resolving the unstable manifold of
$\primeOrb{0}$ enables us to locate these heteroclinic connections
between \po s. Note that 1-dimensional
stable manifold of $\primeOrb{1}$ separates the unstable manifold of
$\primeOrb{0}$ in two pieces. Green and blue orbits in
\reffig{f-Connectionsa} appear to be at two sides of this
invariant boundary: while one of them converges to $\primeOrb{2}$,
the other leaves the neighborhood to explore other parts of the
\statesp\ that are not captured by the Poincar\'e section,
following the unstable manifold of $\primeOrb{1}$.

\begin{figure}[h]
	\centering
	\includegraphics[width=0.9\textwidth]{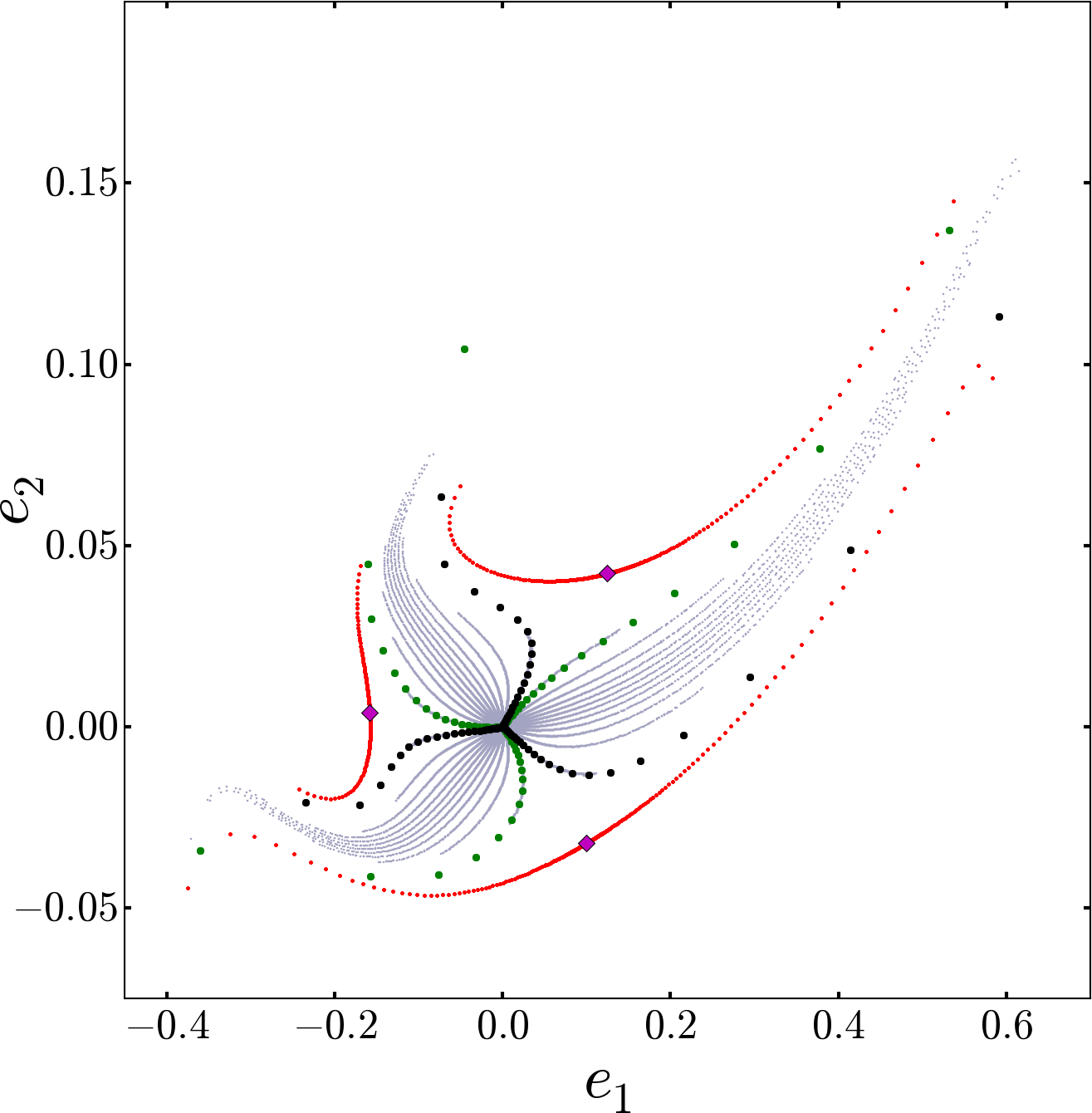}
	\caption{\label{f-Connectionsb}
		Unstable manifold of $\primeOrb{0}$ (gray) and two orbits
		(black and green) within at $L=21.48$. Red points lie on
		the 1-dimensional unstable manifold of $\primeOrb{1}$
		(magenta).
	}		
\end{figure}

As the system size is increased, $\primeOrb{2}$ becomes
unstable at $L \approx 21.38$. At $L \approx 21.477$ the two complex
unstable Floquet multipliers collide on the real axis and at
$L \approx 21.479$ one of them crosses the unit circle.
After this bifurcation, we were no longer able to continue this
orbit. At $L = 21.48$, the spreading of the $\primeOrb{0}$'s
unstable manifold becomes more dramatic, and its boundary is set
by the 1-dimensional unstable manifold of $\primeOrb{1}$, as shown in
\reffig{f-Connectionsb}. We compute the unstable manifold of
$\primeOrb{1}$ similarly to \refeq{e-InitUnstMan}, by integrating
\beq
    \sspRefRed_{\PoincS} (\delta) = \sspRefRed_{\primeOrb{1}, \PoincS} \pm
                          \epsilon \ExpaEig_1^\delta \VRefRed_{1, \PoincS}
    \,, \quad \mbox{where } \delta \in [0, 1) \,.
\label{e-InitUnstMan1D}
\eeq
$\ExpaEig_1$ and $\VRefRed_{1}$ in \refeq{e-InitUnstMan1D} are the
unstable Floquet multiplier and the corresponding Floquet vector
of $\sspRefRed_{\primeOrb{1}}$, and the initial conditions
\refeq{e-InitUnstMan1D} cover the unstable manifold of
$\sspRefRed_{\primeOrb{1}}$ in the linear approximation.

A negative real Floquet multiplier of \primeOrb{1} crosses the unit
circle at $L \approx 21.6$ leading to ``drifting'' dynamics
	in the associated unstable direction.
	Such ``symmetry-breaking'' bifurcations of \rpo s with
	$\Ztwo$ symmetry are ubiquitous in many physical settings:
	Earlier examples are studies of reduced-order models of
    convection\rf{KnoWei81}, forced pendulum\rf{DBHL82}, and
    Duffing oscillator\rf{NoFre82}, which reported that symmetry
    breaking bifurcations precede period doubling route to chaos.
    A key observation was made by Swift and Wiesenfeld\rf{SwiWie84},
    who showed in the context of periodically driven damped pendulum
    that Poincar\'e map associated with the symmetric system is the
    second iterate of another ``reduced'' Poincar\'e map, which
    identifies symmetry-equivalent points. They then argue that
    $\Ztwo$-symmetric periodic orbits generically
    do not undergo period doubling bifurcations when a single
    parameter of the system varied. More recent
    works\rf{MarLopBla04,BlMaLo05} adapt \refref{SwiWie84}'s
    reduced Poincar\'e map to fluid systems in order to study their bifurcations
    in the presence of symmetries. For a review of the
    symmetry-breaking bifurcations in fluid dynamics, see \refref{CraKno91}.

    As in the previous cases, in order
    to investigate the dynamics of the system at this stage, we
	compute and visualize the unstable manifold of \primeOrb{1}.
    
\begin{figure}[h]
	\centering
	\includegraphics[width=0.9\textwidth]{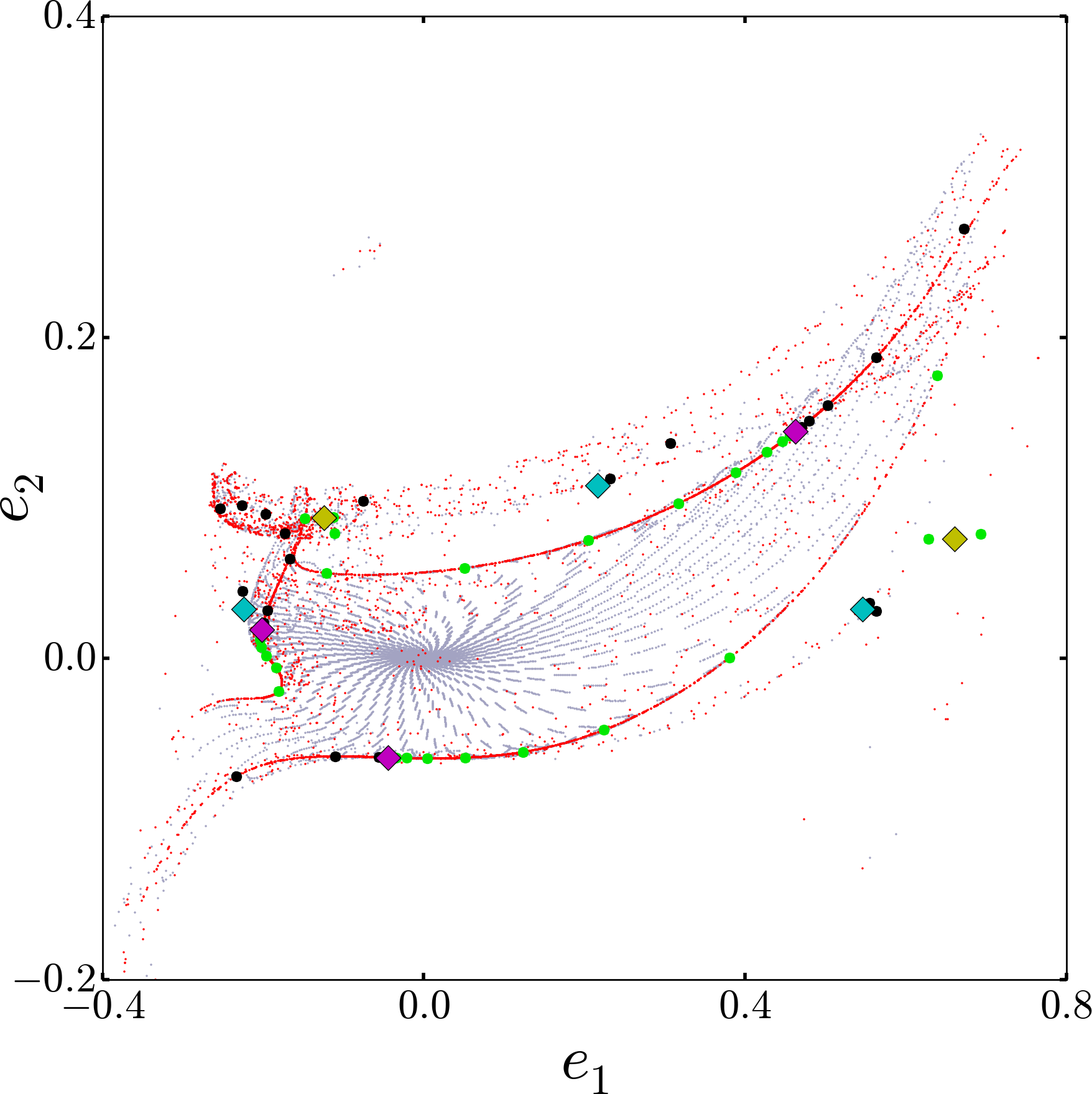}
	\caption{\label{f-ConnectionsL21p7a}
		Unstable manifolds of $\primeOrb{0}$  (gray) and
		$\primeOrb{1}$ (red) on the Poincar\'e
		section \refeq{e-Psect} at $L=21.7$.
		Magenta, cyan, and yellow diamond markers respectively indicate the
		{\PoincSec} points of $\primeOrb{1}$, $\primeOrb{3}$, and
		$\primeOrb{4}$.
		Green and black dots correspond to two individual orbits
		started on the linear approximation to the unstable manifold
		of $\primeOrb{1}$, which visit neighborhoods of
		$\primeOrb{3}$ and $\primeOrb{4}$ respectively.
	}		
\end{figure}

Similarly to
\refeq{e-InitUnstMan} and \refeq{e-InitUnstMan1D}, the 2-dimensional
unstable manifold of \primeOrb{1}
is approximately covered by initial conditions
\beq
		\sspRefRed_{\PoincS} (\delta, \phi) =
		\sspRefRed_{\primeOrb{1}, \PoincS} +
		\epsilon \left[
		|\ExpaEig_1|^\delta \cos \phi \, \VRefRed_{1, \PoincS}
		+ |\ExpaEig_2|^\delta \sin \phi \, \VRefRed_{2, \PoincS}
		\right]
		\label{e-InitUnstMan2DReal}
\eeq
where $\delta \in [0, 1) \, , \; \phi \in [0, 2 \pi)$.
At system size $L=21.7$, we set $\epsilon = 10^{-3}$ and
discretize \refeq{e-InitUnstMan2DReal} by choosing
$10$ and $36$ equally spaced values for $\delta$ and $\phi$,
respectively. First $38$ returns of orbits generated
according to \refeq{e-InitUnstMan2DReal} are shown in
\reffig{f-ConnectionsL21p7a} as red points along with
the unstable manifold of \primeOrb{0} (gray). Note that,
unlike \reffig{f-Connectionsb},
in \reffig{f-ConnectionsL21p7a} there is no clear
separation on the unstable manifold of \primeOrb{0}.
This is because the
connection of \primeOrb{0}'s unstable manifold to
\primeOrb{1} is no longer captured by the {\PoincSec}
\refeq{e-Psect} after the unstable manifold of \primeOrb{1}
becomes 2-dimensional.
Yet, unstable manifold of \primeOrb{1}
still shapes that of \primeOrb{0}.

\begin{figure}[h]
	\floatbox[{\capbeside\thisfloatsetup{capbesideposition={left,center},capbesidewidth=0.4\textwidth}}]{figure}[\FBwidth]
	{\caption{(a) Space-time visualization of a segment of the orbit marked
			green on \reffig{f-ConnectionsL21p7a} as it leaves the neighborhood of $\primeOrb{1}$
			and enters the neighborhood of $\primeOrb{4}$.
			(b) Space-time visualizations of $\primeOrb{1}$ (bottom) and
			$\primeOrb{4}$ (top).}\label{f-ConnectionsL21p7b}}
	{
		\begin{overpic}[height=0.45\textwidth]{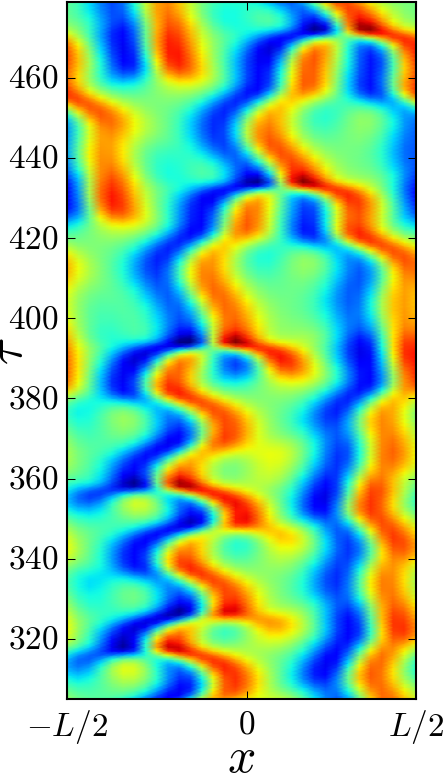}
			\put (0.5,-1) {(a)}
		\end{overpic} \quad
		\begin{overpic}[height=0.45\textwidth]{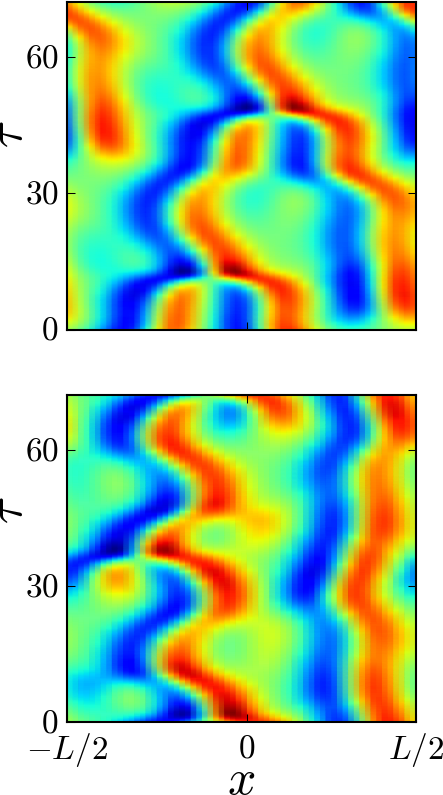}
			\put (0.5,-1) {(b)}
		\end{overpic}
	}
\end{figure}

Since the leading Floquet exponent $\mu_1$ of
\primeOrb{1} is approximately an order of magnitude larger than $\mu_2$
(see \reftab{t-EeigVals}), unstable manifold of \primeOrb{1}
appears as if it is 1-dimensional in \reffig{f-ConnectionsL21p7a}.
However, it is absolutely crucial to study this manifold in $2$
dimensions as different initial conditions in this 2-dimensional
manifold connect to the regions of \statesp\ with qualitatively
different dynamics. In order to illustrate this point, we have
marked two individual trajectories on the unstable manifold of
\primeOrb{1} with black and green in \reffig{f-ConnectionsL21p7a}.
After observing that these orbits have nearly recurrent dynamics,
we ran Newton searches in their vicinity and found two new
periodic orbits $\primeOrb{3}$ and $\primeOrb{4}$,
marked respectively with cyan and yellow diamonds
on \reffig{f-ConnectionsL21p7a}. In the full \statesp\
$\primeOrb{3}$ is a pre-\po\
\refeq{e-PPO}, whereas $\primeOrb{4}$ is a \rpo\ \refeq{e-RPO}
with a non-zero drift. We show a time segment of the
orbit marked green on \reffig{f-ConnectionsL21p7a}
without symmetry reduction, as
color-coded amplitude of the scalar field $u(\conf, \zeit)$ in
\reffig{f-ConnectionsL21p7b}\,(a).
For comparison we also show two repeats of
\primeOrb{1} (bottom) and \primeOrb{4} (top) in
\reffig{f-ConnectionsL21p7b}\,(b). \refFig{f-ConnectionsL21p7b}
suggests that this orbit leaves the neighborhood of
\primeOrb{1} following a heteroclinic connection to \primeOrb{4}.

In \reffig{f-ConnectionsL21p7a}, some of the red points appear on
the unstable manifold of \primeOrb{0}. These points corresponds to
trajectories that leave the unstable manifold of \primeOrb{1}, come back
after exploring other parts of the \statesp\ and follow unstable manifold
of \primeOrb{0}. We could have excluded these points by showing shorter
trajectories for higher values of $\delta$ in \refeq{e-InitUnstMan2DReal}
in \reffig{f-ConnectionsL21p7a}, however we chose not to do so
in order to stress that visualizations of unstable manifolds of \po s
are not restricted to the dynamics within a small neighborhood of a \po,
but in fact they illuminate the geometry of the flow in a finite part of
the strange attractor.

An interesting feature of the bifurcation scenario studied here is the
apparent destabilization of the invariant torus before its breakdown.
Note that in \reffig{f-UnstMan21p31a} the trajectories within
the unstable manifold of \primeOrb{0} diverge in normal direction from the
region that was inhabited by a stable 2-torus for lower values of $L$. This
suggest that the invariant torus has become normally
hyperbolic\rf{Feniche71}.
This torus could be computed by the method of \refref{LCC06}, but our
goal here is more modest, what we have computed already amply
demonstrates the utility of our \On{2} symmetry reduction.
Note also that the
stable period-3 orbit $\primeOrb{2}$ in \reffig{f-Connectionsa} has a finite basin
of attraction, and the trajectories which do not fall into it leave its
neighborhood. In typical scenarios involving generation of stable
- unstable pairs of periodic orbits within an invariant torus (see \eg\
\refref{arnold82}), the torus becomes a heteroclinic connection between the
periodic orbit pair. Here the birth of the period-3 orbits appears to
destroy the torus.

\section{Summary and future directions}
\label{s-Discuss}

The two main results presented here are:
1) a new method for reducing the \On{2}-symmetry of PDEs,
and
2) a symmetry-reduced \statesp\ \Poincare\ section visualization of
1- and 2-dimensional unstable manifolds of \KS\ \po s.

Our method for the computation of unstable manifolds is
general and can find applications in many other ODE and PDE settings.
The main idea here is a generalization of Gibson \etal\rf{GHCW07} method
for visualizations of the unstable manifolds of \eqva, originally applied
to plane Couette flow, a setting much more complex then the current
paper. All our computations are carried
out for the full \KSe\ \refeq{e-ks}, in 30 dimensions, and it is
remarkable how much information is captured by the 2- and 3-dimensional
projections of the \On{2} symmetry-reduced {\PoincSec s} - none of
that structure is visible in the full \statesp.

The \KS\ \On{2} symmetry reduction method described here might require
modifications when applied to other problems. For example,
for PDEs of space dimensions larger then one,
there can be more freedom in choosing the phase fixing
condition \refeq{e-fFslice}.
	This indeed is the case for shear flows with both
	homogeneous (streamwise and spanwise translation invariant) and inhomogeneous
	(wall-normal) directions. When adapting the \fFslice\ method
	to such problems, one should experiment with the dependence
	of the phase fixing condition on the inhomogeneous coordinate
	such that the slice fixing phase is uniquely defined for
	\statesp\ regions of interest; see chapter 3 of
	\refref{BudanurThesis} for details.
\refRef{WiShCv15} makes this choice for pipe
flow by taking a `typical state' in the turbulent flow, setting all
streamwise Fourier modes other than the first one to zero, and using
this state as a ``slice template''.
	Another point to be taken into
	consideration for canonical shear flows is that their symmetry
	group 	is $\SOn{2} \times \On{2}$. So far, continuous
	symmetry reduction in pipe flows\rf{ACHKW11,WiShCv15}
	were confined to settings, where an imposed symmetry in conjugacy
	class of spanwise reflection disallows spanwise rotations.
	When no such restriction is present, one needs two
	conditions for fixing both streamwise and spanwise translations.
	These conditions must be chosen such that the order at which
	continuous symmetries are reduced does not matter. For
	direct products of commuting $\SOn{2}$ symmetries, this is
	a straightforward task and outlined in section 3 of
	\refref{BudanurThesis}. An 	application of these ideas to the
	pipe flow is going to appear in a future publication\rf{BudHof17}.

Furthermore, while invariant polynomials similar to (\ref{e-RefInvs}) can
be constructed for any problem with a reflection symmetry, an intermediate
step is necessary if the action of reflection $\sigma$ symmetry is not
the sign flip of a subset of coordinates. In that case, one should first
decompose the \statesp\ into symmetric and antisymmetric subspaces by
computing
$\ssp_S = (1/2) [\ssp + \matrixRep (\sigma) \ssp]$ and
$\ssp_A = (1/2) [\ssp - \matrixRep (\sigma) \ssp]$, respectively,
and construct invariants analogous to (\ref{e-RefInvs}) for elements of
$\ssp_A$ that are not strictly zero.
Generalizations of this approach to richer discrete symmetries, such
as dihedral groups, remains an open problem, with potential
application to systems such as the Kolmogorov
flow\rf{PlaSirFit91,Faraz15}.

Bifurcation scenarios similar to the one studied here are
ubiquitous in high-dimensional systems. For example, Avila
\etal\rf{AvMeRoHo13} study of transition to turbulence in pipe flow, and
Zammert and Eckhardt's study of the plane Poiseuille flow\rf{ZamEck15}
both report torus bifurcations of \rpo s along transitions to chaos. We
believe that the methods presented in this paper can lead to a
deeper understanding of these scenarios.

While unstable manifold visualizations of \po s in the
symmetry-reduced \statesp\ illustrates bifurcations of these orbits,
our motivation for
investigating such objects is not a study of bifurcations, but ultimately
a partition of the turbulent flow's
\statesp\ into qualitatively different regions, and
construction of the corresponding symbolic dynamics.
\refFig{f-ConnectionsL21p7a} and \reffig{f-ConnectionsL21p7b} demonstrate
our progress in this direction:
we are able to identify symmetry breaking heteroclinic connections from
non-drifting solutions to the drifting ones.
Such observations would have been very hard to make without reducing
symmetries of the system, since each \rpo\ has a reflection copy,
corresponding to a solution drifting in the other direction; and each
such solution has infinitely many copies obtained by translations.

\begin{acknowledgements}
We are grateful to
Xiong Ding,
Evangelos Siminos,
Simon Berman,
and
Mohammad Farazmand
for many fruitful discussions.
\end{acknowledgements}

\appendix
\section{Computational details}
\label{s-Stability}

Throughout this paper, we used the $16$ Fourier mode truncation of
\KSe\ \refeq{e-Fks}, which renders the \statesp\ $30$-dimensional.
Sufficiency of this truncation was
demonstrated for $L=22$ in \refref{SCD07}. In all our computations, we
integrate \refeq{e-velRed} and its gradient system numerically,
using a general purpose adaptive integrator \texttt{odeint} from
\texttt{scipy.integrate}\rf{scipy}, which is a wrapper of
\texttt{lsoda} from \texttt{ODEPACK} library\rf{hindmarsh1983}.
Note that \refeq{e-velRed} is singular if $\hat{b}_1 = 0$, \ie,
whenever the first Fourier mode vanishes.
This singularity can be regularized by a time-rescaling if a fixed
time step integrator is desired\rf{BudCvi14}.

Transformation of trajectories and tangent vectors to the fully
symmetry-reduced \statesp\ \refeq{e-sspRefRed} is applied as
post-processing. For a trajectory $\sspRed (\zeit)$, we simply apply
the reflection reducing transformation to obtain the trajectory as
$\sspRefRed(\zeit) = \sspRefRed(\sspRed (\zeit)) $. Velocity field
\refeq{e-velRed} transforms to \refeq{e-sspRefRed} by acting with
the \jacobianM
\[
    \velRefRed(\sspRefRed) =
    \frac{d \sspRefRed(\sspRed)}{d \sspRed}\velRed(\sspRed)
\,.
\]

Floquet vectors transform to the fully symmetry-reduced \statesp\ similarly,
however, their computations in the \fFslice\ requires some care.
Remember that the reflection symmetry remains after the continuous
symmetry reduction, and its action is represented by (\ref{e-DRRed}).
Thus, denoting finite time flow induced by \refeq{e-velRed} by
$\flowRed{\zeit}{\sspRed}$, pre-\po\  within the \slice\ satisfies
\[
\sspRedPPO = \matrixRepRed (\sigma)
\flowRed{\period{p}}{\sspRedPPO} \,,
\]
with its linear stability given by the spectrum of the \jacobianM
\[
\jMpsRed_{\!_{pp}} = \matrixRepRed (\sigma)
\jMpsRed^{\period{p}} (\sspRedPPO) \,,
\]
where $\jMpsRed^{\period{p}} (\sspRedPPO)$ is the \jacobianM\ of the
flow function $\flowRed{\period{p}}{\sspRedPPO}$. Thus, in order to
find the Floquet vectors in fully symmetry-reduced representation, we first
find the eigenvectors $\VRed$ of the \jacobianM\ $\jMpsRed_{\!_{pp}}$
and then transform them as
\(
\VRefRed(\sspRefRed) =
{d \sspRefRed(\sspRedPPO)}/{d \sspRed}\,\VRed(\sspRed)
\,.
\)

\bibliographystyle{spphys}       
\bibliography{../siminos/bibtex/siminos}


\end{document}